\pdfoutput=1
\documentclass[nohyperref]{lecturenotes}

\InputIfFileExists{hg.id}{}{\def\HgNode{??}\def\HgDate{??}}
\def\HgNode{7b021525c2d7}\def\HgDate{2012-07-17 09:55 +0100}

\title{Managing Research Data in Big Science}
\author{Norman Gray, Tobia Carozzi and Graham Woan\strut\\
  \strut
  SUPA School of Physics and Astronomy,
  University of Glasgow}

\newif\ifsnapshot 
\snapshotfalse
\def\documentdate{2011 July 14}
\def\documentident{v1.1}

\ifsnapshot
\usepackage{color}
\definecolor{draftgrey}{gray}{0.75}
\fi

\usepackage{recommendations}

\usepackage{prettyref}
\newrefformat{s}{Sect.~\ref{#1}}
\usepackage{varioref}
\newrefformat{f}{Fig.~\vref{#1}}
\newrefformat{tab}{Table~\vref{#1}}

\usepackage[xindy]{glossaries}

\usepackage[load-configurations=binary]{siunitx}
\DeclareSIUnit{\yr}{yr}
\DeclareSIUnit{\mth}{month}           
\ifxetex
  \DeclareSIUnit{\eur}{€}
\else
  \makeatletter
  \newbox\fakeeurobox
  \setbox\fakeeurobox=\hbox{%
    \valign{#\cr
      \vfil\hbox to 0.07em{\dimen@\f@size\p@
                           \math@fontsfalse
                           \fontsize{.7\dimen@}\z@\selectfont=\hss}\vfil\cr%
       \hbox{C}\crcr
    }%
  }
  \makeatother
  \def\fakeeuro{\copy\fakeeurobox}
  \DeclareSIUnit{\eur}{\fakeeuro }
\fi
\DeclareSIUnit{\PBY}{\peta\byte\per\yr}

\ifxetex
  \def\q#1{‘#1’}
  \def\qq#1{“#1”}
  \def\dash{~–\space}
  \def\range{–}
  \def\ccedilla{ç}
  \def\oacute{ó}
  \newfontface\smallcapsfont[Scale=0.8,LetterSpace=3.0]{Optima}
  \renewcommand\textsc[1]{{\smallcapsfont \uppercase{#1}}}
  \PassOptionsToPackage{xetex,unicode}{hyperref}
\else
  \def\q#1{`#1'}
  \def\qq#1{``#1''}
  \def\dash{~--\space}
  \def\range{--}
  \def\ccedilla{{\c c}}
  \def\oacute{{\'o}}
  \usepackage[scaled=0.92]{helvet} 
  \usepackage{microtype}
  \PassOptionsToPackage{pdftex}{hyperref}
\fi

\usepackage[strict]{changepage} 

\newif\ifendnotes
\endnotesfalse
\ifendnotes
  \usepackage{endnotes}
  \let\footnote\endnote
\else
  \usepackage{marginfix}
  \makeatletter\let\MFX@debug\@gobble\makeatother 
  \def\marginnote#1{\marginpar[{\raggedright\fontsize{7}{9}\selectfont #1}]{\raggedleft\fontsize{7}{9}\selectfont #1}}
  \let\footnote\marginnote
\fi

\usepackage[hyperindex=false,pdftitle={Managing Research Data in Big Science},pdfauthor={Norman Gray, Tobia Carozzi and Graham Woan}]{hyperref}

\ifxetex
  \def\UrlFont{\ttfamily
    \dimen0=1ex
    \ifdim\dimen0<4pt
      \fontsize{7pt}{8pt}
    \else
      \fontsize{9pt}{10pt}%
    \fi
    \selectfont}
\else
  \def\UrlFont{\ttfamily
    \dimen0=1ex
    \ifdim\dimen0<4pt
      \fontsize{6pt}{8pt}
    \else
      \fontsize{8pt}{10pt}%
    \fi
    \selectfont}
\fi

\def\adhoc{\emph{ad hoc}}

\def\longurlX#1{\UrlFont #1\endgroup}
\def\longurl{\begingroup \def~{\discretionary{}{}{}} \longurlX}

\CopyrightStatement{Copyright 2011\range2012, University of Glasgow}

\tocstyle{noleaders}

\newif\ifcasestudy
\casestudyfalse

\date{\documentdate, \documentident}

\newglossaryentry{ATLAS}{name={ATLAS},description={One of the four detectors at the LHC, and one of the two large ones}}
\newglossaryentry{CMS}{name={CMS},description={Compact Muon Solenoid: One of the four detectors at the LHC, and one of the two large ones},first={Compact Muon Solenoid (CMS)},firstplural={Compact Muon Solenoids (CMSs)}}
\newglossaryentry{DCC}{name={DCC},description={Digital Curation Centre: \url{http://www.dcc.ac.uk} (not to be confused with the LSC Document Control Center)},first={Digital Curation Centre (DCC)},firstplural={Digital Curation Centres (DCCs)}}
\newglossaryentry{LHC}{name={LHC},description={The Large Hadron Collider at CERN: the accelerator is the host for two large general purpose detectors (ATLAS and CMS) and two smaller ones (ALICE and LHCb)}}
\newglossaryentry{NSF}{name={NSF},description={National Science Foundation: the principal (non-defence) science funder in the USA},first={National Science Foundation (NSF)},firstplural={National Science Foundations (NSFs)}}
\newglossaryentry{STFC}{name={STFC},description={Science and Technology Facilities Council: the primary UK funder of facility-scale science, see \url{http://www.stfc.ac.uk}},first={Science and Technology Facilities Council (STFC)},firstplural={Science and Technology Facilities Councils (STFCs)}}
\newglossaryentry{ESRC}{name={ESRC},description={Economic and Social Research Council: the principal social science funder in the UK, see \url{http://www.esrc.ac.uk}},first={Economic and Social Research Council (ESRC)},firstplural={Economic and Social Research Councils (ESRCs)}}
\newglossaryentry{JISC}{name={JISC},description={Joint Information Systems Committee: The organisation responsible for the maintenance and effective exploitation of the academic computing network in the UK, and the funders of this present report},first={Joint Information Systems Committee (JISC)},firstplural={Joint Information Systems Committees (JISCs)}}
\newglossaryentry{DASWG}{name={DASWG},description={Data Analysis and Software Working Group: the principal LIGO software forum},first={Data Analysis and Software Working Group (DASWG)},firstplural={Data Analysis and Software Working Groups (DASWGs)}}
\newglossaryentry{IVOA}{name={IVOA},description={International Virtual Observatory Alliance: the consortium which defines VO standards},first={International Virtual Observatory Alliance (IVOA)},firstplural={International Virtual Observatory Alliances (IVOAs)}}
\newglossaryentry{VO}{name={VO},description={Virtual Observatory: a set of data sharing argreements and protocols.  See \prettyref{s:vo} (not to be confused with grid Virtual Organisations)},first={Virtual Observatory (VO)},firstplural={Virtual Observatorys (VOs)}}
\newglossaryentry{VOEvents}{name={VOEvents},description={The components of an alerting system for astronomical events, standardised by the IVOA}}
\newglossaryentry{strain data}{name={strain data},description={The fundamental GW signal}}
\newglossaryentry{SKA}{name={SKA},description={Square Kilometre Array: a low frequency radio telescope with a large (one square kilometre) collecting area},first={Square Kilometre Array (SKA)},firstplural={Square Kilometre Arrays (SKAs)}}
\newglossaryentry{data sharing}{name={data sharing},description={The formalised practice of making science data publicly available}}
\newacronym{DMP}{DMP}{Data Management and Preservation}
\newglossaryentry{FITS}{name={FITS},description={Flexible Image Transport System: the standard file format in astronomy, see \url{http://fits.gsfc.nasa.gov}},first={Flexible Image Transport System (FITS)},firstplural={Flexible Image Transport Systems (FITSs)}}
\newglossaryentry{data products}{name={data products},description={Formal data outputs from an observatory, instrument or process (see \prettyref{s:rawdata})}}
\newglossaryentry{pipeline}{name={pipeline},description={A software system (or sometimes a software-hardware hybrid) which transforms raw data into more or more levels of data product.  The data reduction pipelines, which must be able to keep up with the rate at which data is acquired, and which are assembled from a mixture of standard and custom software components, generally absorb a significant fraction of the total development budget of a new instrument}}
\newglossaryentry{raw data}{name={raw data},description={The data extracted directly from an instrument or observation; since it is uncalibrated and uncorrected, it is generally of little use to those not intimately familiar with the instrument (see \prettyref{s:rawdata})}}
\newacronym{HEP}{HEP}{High Energy Physics}
\newglossaryentry{catalogue}{name={catalogue},description={In the astronomical context, a catalogue is a table of positions and other information for stars or other other astronomical objects}}
\newglossaryentry{KRDS}{name={KRDS},description={Keeping Research Data Safe: JISC project developing and documenting data preservation tools and studies; see \url{http://www.beagrie.com/krds.php} and \cite{beagrie10}},first={Keeping Research Data Safe (KRDS)},firstplural={Keeping Research Data Safes (KRDSs)}}
\newglossaryentry{LIGO}{name={LIGO},description={Laser Interferometer Gravitational-wave Observatory: the hardware, comprising LIGO Lab and GEO (see \url{http://ligo.org} and \prettyref{s:consortia})}}
\newglossaryentry{aLIGO}{name={aLIGO},description={Advanced LIGO: The successor project to LIGO, due to start in 2015},first={Advanced LIGO (aLIGO)},firstplural={Advanced LIGOs (aLIGOs)}}
\newglossaryentry{LIGO Lab}{name={LIGO Lab},description={The Caltech/MIT consortium, funded by NSF to design and run the LIGO interferometers in the US, see \url{http://www.ligo.caltech.edu}}}
\newglossaryentry{LSC}{name={LSC},description={LIGO Scientific Collaboration: The network of research groups contributing effort to the LIGO experiment and data analysis, see \url{http://ligo.org}},first={LIGO Scientific Collaboration (LSC)},firstplural={LIGO Scientific Collaborations (LSCs)}}
\newglossaryentry{LVC}{name={LVC},description={A data-sharing agreement between the LSC and the Virgo Collaboration (see \prettyref{s:consortia})}}
\newglossaryentry{Virgo}{name={Virgo},description={Italian-French gravitational-wave detector \url{http://www.virgo.infn.it/}}}
\newglossaryentry{GEO}{name={GEO},description={A German-British consortium, responsible for the GEO600 interferometer, funded jointly by STFC and the German government}}
\newglossaryentry{GEO600}{name={GEO600},description={The GEO observatory located near Hannover in Germany}}
\newglossaryentry{RDS}{name={RDS},description={Reduced data sets: the core hierarchy of LIGO data sets},first={Reduced data sets (RDS)},firstplural={Reduced data setss (RDSs)}}
\newacronym{GW}{GW}{Gravitational Wave}
\newglossaryentry{big science}{name={big science},description={A class of science projects characterised by being international, highly collaborative and expensively funded (see \prettyref{s:bigscience} for more discussion)}}
\newglossaryentry{HERA}{name={HERA},description={A particle accelerator at the German DESY facility}}
\newglossaryentry{MRD}{name={MRD},description={Managing Research Data: A funding programme within the JISC e-Research theme, see \url{http://www.jisc.ac.uk/whatwedo/programmes/mrd}},first={Managing Research Data (MRD)},firstplural={Managing Research Datas (MRDs)}}
\newglossaryentry{CCSDS}{name={CCSDS},description={Consultative Committee for Space Data Systems: authors of the OAIS reference model, see \url{http://www.ccsds.org}},first={Consultative Committee for Space Data Systems (CCSDS)},firstplural={Consultative Committee for Space Data Systemss (CCSDSs)}}
\newglossaryentry{MOU}{name={MOU},description={Memorandum of Understanding: the relationships between the various participating entities and the LSC is articulated through a series of annually reviewed MOUs},first={Memorandum of Understanding (MOU)},firstplural={Memorandum of Understandings (MOUs)}}
\newglossaryentry{CDS}{name={CDS},description={Strasbourg Data Centre: (see \url{http://cdsweb.u-strasbg.fr/} and \prettyref{s:cds})},first={Strasbourg Data Centre (CDS)},firstplural={Strasbourg Data Centres (CDSs)}}
\newglossaryentry{ADS}{name={ADS},description={Astrophysical Data Service: a bibliographic archive for astronomy, based at the Harvard-Smithsonian Center for Astrophysics; ADS preserves full-text copies of journal articles, both in collaboration with publishers, and through a digitization process, and maintains a widely-used bibliographic ID system (\url{http://ads.harvard.edu})},first={Astrophysical Data Service (ADS)},firstplural={Astrophysical Data Services (ADSs)}}
\newglossaryentry{VizieR}{name={VizieR},description={A repository of astronomical catalogue data at CDS (see \prettyref{s:cds})}}
\newglossaryentry{arXiv}{name={arXiv},description={A electronic preprint service, see \url{http://arxiv.org}.  The arXiv started in the early 90s, based on FTP and email.  It initially serviced particle physics and astronomy, but has expanded to cover other areas of physics, mathematics, and some areas of computing science}}
\newglossaryentry{NSSDC}{name={NSSDC},description={National Space Science Data Center: the permanent archive for NASA space science mission data \url{http://nssdc.gsfc.nasa.gov}},first={National Space Science Data Center (NSSDC)},firstplural={National Space Science Data Centers (NSSDCs)}}
\newglossaryentry{PDS}{name={PDS},description={Planetary Data System: The NASA data archive and standard set \url{http://pds.nasa.gov/}},first={Planetary Data System (PDS)},firstplural={Planetary Data Systems (PDSs)}}
\newglossaryentry{NASA}{name={NASA},description={National Aeronautics and Space Administration: The US space agency \url{http://www.nasa.gov}}}
\newglossaryentry{ESA}{name={ESA},description={European Space Agency: \url{http://www.esa.int}},first={European Space Agency (ESA)},firstplural={European Space Agencys (ESAs)}}
\newglossaryentry{ESO}{name={ESO},description={European Southern Observatory: A pan-european agency running a set of southern-hemisphere telescopes \url{http://www.eso.org}},first={European Southern Observatory (ESO)},firstplural={European Southern Observatorys (ESOs)}}
\newglossaryentry{Data Object}{name={Data Object},description={Either a Physical Object or a Digital Object (OAIS) (that is, the `Data Object' is the sequence of bits, or the physical object which is \emph{the data} in the most primitive sense)}}
\newglossaryentry{Designated Community}{name={Designated Community},description={An identified group of potential Consumers who should be able to understand a particular set of information (OAIS)}}
\newglossaryentry{Representation Information}{name={Representation Information},description={The information that maps a Data Object into more meaningful concepts (OAIS)}}
\newglossaryentry{Representation Network}{name={Representation Network},description={The set of Representation Information that fully describes the meaning of a Data Object. Representation Information in digital forms needs additional Representation Information so its digital forms can be understood over the Long Term (OAIS)}}
\newglossaryentry{Content Information}{name={Content Information},description={The set of information that is the original target of preservation by the OAIS (OAIS)}}
\newglossaryentry{SIP}{name={SIP},description={Submission Information Package: An Information Package that is delivered by the Producer to the OAIS for use in the construction of one or more AIPs (OAIS)},first={Submission Information Package (SIP)},firstplural={Submission Information Packages (SIPs)}}
\newglossaryentry{AIP}{name={AIP},description={Archival Information Package: An Information Package, consisting of the Content Information and the associated {Preservation Description Information}, which is preserved within an OAIS (OAIS)},first={Archival Information Package (AIP)},firstplural={Archival Information Packages (AIPs)}}
\newglossaryentry{DIP}{name={DIP},description={Dissemination Information Package: The Information Package, derived from one or more AIPs, received by the Consumer in response to a request to the OAIS (OAIS)},first={Dissemination Information Package (DIP)},firstplural={Dissemination Information Packages (DIPs)}}
\newglossaryentry{Preservation Description Information}{name={Preservation Description Information},description={The information which is necessary for adequate preservation of the Content Information and which can be categorized as Provenance, Reference, Fixity, and Context information (OAIS)}}
\newglossaryentry{Long Term}{name={Long Term},description={A period of time long enough for there to be concern about the impacts of changing technologies, including support for new media and data formats, and of a changing user community, on the information being held in a repository. This period extends into the indefinite future (OAIS)}}
\newglossaryentry{Information Package}{name={Information Package},description={The Content Information and associated Preservation Description Information which is needed to aid in the preservation of the Content Information. The Information Package has associated Packaging Information used to delimit and identify the Content Information and Preservation Description Information (OAIS)}}

\makeindex

\begin{document}

\maketitle
\fancyhead[el]{Gray, Carozzi and Woan} 

\medskip
\noindent
\begin{centering}
\def\arraystretch{1.3}
\begin{tabular}{|rl|}
\hline
Version &\documentident\\
\ifsnapshot
  \textbf{SNAPSHOT} & \HgNode, \HgDate \\
\fi
URL& \url{http://purl.org/nxg/projects/mrd-gw/report}\\
Distribution& Public\\
\hline
\end{tabular}\\
\end{centering}
\medskip

\noindent
This report was prepared as part of the RDMP strand of the JISC programme
\href{http://www.jisc.ac.uk/whatwedo/programmes/mrd}{Managing Research Data}.

\section*{Abstract}

The project which led to this report was funded by JISC in 2010\range2011
as part of its \q{Managing Research Data} programme, to examine the
way in which Big Science data is managed, and produce any
recommendations which may be appropriate.

Big science data is different: it comes in large volumes, and it is
shared and exploited in ways which may differ from other
disciplines. This project has explored these differences using as a
case-study Gravitational Wave data generated by the LSC, and has
produced recommendations intended to be useful
variously to JISC, the funding council (STFC) and the LSC community.

In \prettyref{s:education} we define what we mean by \q{big science},
describe the overall data culture there, laying stress on how it
necessarily or contingently differs from other disciplines.

In \prettyref{s:responsibilities} we discuss the benefits of a formal
data-preservation strategy, and the cases for open data and for
well-preserved data that follow from that.  This leads to our
recommendations that, in essence, funders should adopt rather
light-touch prescriptions regarding data preservation planning: normal
data management practice, in the areas under study, corresponds to
notably good practice in most other areas, so that the only change we
suggest is to make this planning more formal, which makes it more
easily auditable, and more amenable to constructive criticism.

In \prettyref{s:practicalities} we briefly discuss the
LIGO data management plan, and pull together whatever
information is available on the estimation of digital preservation costs.

The report is informed, throughout, by the OAIS reference model for an
open archive.
Some of the report's findings and conclusions were summarised
in~\cite{gray11b}.

See the document history on page~\pageref{s:documenthistory}.

\newpage

\thispagestyle{empty}
\begingroup
\parindent=0pt
\parskip=\medskipamount

\null
\vfill
\includegraphics[height=10mm]{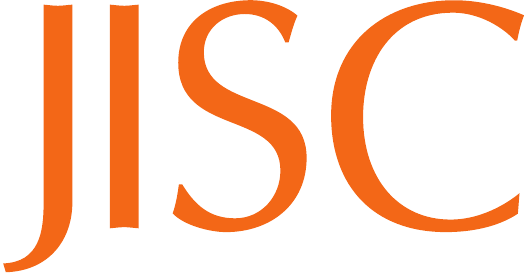}\\[\bigskipamount]
This report was prepared for, and funded by, the RDMP strand of the JISC programme
\href{http://www.jisc.ac.uk/whatwedo/programmes/mrd}{Managing Research Data}.

Release: \HgNode, \HgDate

\makeatletter
\@CopyrightStatement.
\makeatother
This work is licensed under the Creative Commons
   Attribution-Share Alike 2.5 UK: Scotland
   Licence. To view a copy of this licence, visit
   \url{http://creativecommons.org/licenses/by-sa/2.5/scotland/}.
\endgroup
\newpage

\tableofcontents
\makeglossaries

\clearpage

\setcounter{section}{-1}
\section{Introduction}

Astronomy is as old as human culture.  Early agricultural
civilisations required reliable predictions of the positions and
motions of the Sun and Moon, in order to predict in turn seasons,
tides, and river risings.  Even in the absence of an extensive
scientific model, these predictions relied on careful observations,
preserved in the form of almanacs or ephemerides.
Documents such as these associate astronomy
with not only the first data archives but, since these artifacts
still exist, also the oldest data archives in the world.
Long-term digital preservation in astronomy is nothing new.  We cannot
resist saying more about this, in \prettyref{s:babylon}.

Astronomical archiving does however evolve, and in the last few
decades both astronomy and particle physics have had to become leaders
in large-scale data management.

Although astronomical images (now all born digital) have always been
substantial in size, they have generally been reasonably manageable.
Newer astronomical techniques\dash and we are thinking of 21st century
radio astronomy and gravitational astronomy\dash are capable of generating
truly challenging quantities of data; and particle physics has been
generating, and addressing, intimidating data problems for decades.
These problems cover both the management and preservation of large data
volumes, as technical problems, and the preservation of the data's
information content, on substantially varying timescales.


\subsection{Project Background}
The Managing Research Data/Gravitational Waves project (MRD-GW) is
concerned with the data management arrangements of the \gls{LSC},
and of the broader \gls{GW} community.  It is one of the six projects
in the RDMP strand of the JISC \gls{MRD}
programme~\cite{jisc-mrd-programme}.

The GW community was selected by the \gls{STFC}, at JISC's invitation, as a
representative example of big-science data management practice\dash as we
elaborate below, it has features of both the traditional astronomy and
HEP communities, without being identifiable with either of them.  While many of
the specifics, below, relate to this community, we believe much of the discussion is relevant to the
other disciplines.
Here\label{s:curatingbigscience}, we are focusing on the big-science projects which
receive strategic support from \gls{STFC}, rather than the
smaller-scale projects funded by specific research grants, since it is
these large-scale projects that are distinctive about STFC-funded
research.  We assume that the outputs of the smaller projects will be
managed through disciplinary repositories, in a manner which more
closely resembles that of other research councils.

The MRD-GW project exists to inform three sets of stakeholders:
\begin{itemize}
\item Although the \gls{JISC} and the \gls{DCC} have extensive experience with digital
  libraries and digital curation in general, there are problems
  specific to \q{\gls{big science}} data which JISC would like to better
  understand.
\item The Research Councils have recently started to require bidders
  to include a \q{data management plan} within project proposals.
  However there is no consensus on what such a plan should look like
  for science funded by the \gls{STFC}.  The US \gls{NSF} has recently
  placed binding requirements on projects to produce data management
  plans~\cite{nsf11}.
\item The LSC community has considerable internal software and
  administration experience, and has solved a large number of data
  management problems focused on large-scale data storage and
  transport.  However there is an awareness that (partly because there
  have been no immediate imperatives to do so) there was until recently no published
  plan for a long-term data archive.
\end{itemize}
The existence of these three groups is reflected in the overall
structure of the document.

This project's context also includes the broad \gls{VO} movement,
which aims to develop standards and areas of consensus which help scientists
have ready access to astronomical data across sub-disciplines and
wavelengths.  All the stakeholder groups have interests in the
success of the \gls{VO} movement.

The project aims to bring together two sets
of practice, namely the long-term digital preservation perspectives
represented by the OAIS reference model\index{OAIS} in the abstract and the \gls{DCC} in
particular, and the very considerable experience of practical
large-scale data management, embedded within the LSC community.\footnote{For
OAIS, see~\cite{std:oais} and \prettyref{s:oais}; in this report the \q{DCC} is the
JISC Digital Curation Centre, not the LIGO Document Control Center.}


\subsection{How to read this document}
\label{s:howtoread}

This document is organised into three main sections, broadly
corresponding to the three audiences we are addressing.

\prettyref{s:education} is about data management in \gls{big science}.
It is addressed to the \gls{JISC} and to the data preservation
community in general, and is intended to illuminate the ways in which
scientists in these areas have distinctive data management
requirements, and a distinctive data culture, which contrasts
informatively with other disciplines.

\prettyref{s:responsibilities} is primarily addressed to \gls{STFC} and
other similar funders of this type of science.  It is concerned with
the responsibilities which are imposed on funders by the wider society,
and which are passed on to the funded through requirements on the
governance of projects and the availability of data.  The
recommendations here are concerned with how best to express these
responsibilities.

Finally, \prettyref{s:practicalities} is primarily addressed to the
\gls{LSC}, as a proxy for similar big-science projects.  The explicit
recommendations here are intended to be of as
much interest to projects, as actions they may wish to take, as to
funders, as behaviour it may be prudent or productive to require.

\subsection{Working with communities\dash pragmatics}

This report is the result of a fruitful collaboration with the
\gls{GW} community.  It may be useful to note some of the features of
the project, and the community, which contributed to this.
\begin{itemize}
\item The project team, as part of Glasgow University, has current
  involvement in the community, and the project director (Woan) is a senior figure there.
\item The \gls{LIGO} community is already aware of the general need for data
  management, and the specific need for preservation (see~\cite{anderson11}).
\item The project personnel have relevant scientific background, and
  are to some extent in the position of being
  informatics-for-astronomy specialists (ie we're \q{insiders}).
\item The community is large and (via studies such
  as~\cite{collins04}) has some experience of being \q{studied}.
\item The existing \gls{LVC} workshop series meant that we could contact
  relevant people easily in a context where newcomers were expected,
  and we didn't have to add our own data management workshop.
\end{itemize}



\section{Data management in Big Science}
\label{s:education}


\subsection{LIGO in perspective: LIGO, big science, and astronomy}
\label{s:bigscience}

What is \q{\gls{big science}}?
\index{big science}

Big science projects tend to share many features which distinguish
them from the way that experimental science has worked in the past.
Such projects share (non-independent) features such as:
\begin{description}
\item[big discoveries] These projects are expected to be amongst the
  most important ones of their generation.  Although there is very
  high confidence that their headline science goals (for example the Higgs and
  GW searches) will be successful, they are also expected to produce long lists
  of unexpected results, and a broad range of engineering spinoffs.
\item[big money] These are decades-long projects, supported by
  country-scale funders and billion-Euro budgets (the total budget for
  the \gls{LHC} is around three billion
  Euros\footnote{\url{http://askanexpert.web.cern.ch/AskAnExpert/en/Accelerators/LHCgeneral-en.html}},
  not including the detectors, nor the personnel and hardware costs
  directly supported by country funders, which cost between one and
  two times that sum).
\item[big author lists] The projects involve collaborations of hundreds of people (the LSC
  author list runs at around 600 people (cf~\prettyref{s:consortia}), and the LHC's \gls{ATLAS} detector
  author list is around 3000).
\item[big data] Enhanced- and Advanced-LIGO (for example) will
  produce of order \SI{1}{\PBY}, comparable to the
  \gls{ATLAS} detector's \SI{10}{\PBY}; the eventual SKA data volumes
  will dwarf these.
\item[big admin] MOUs, councils, workshop series.
\item[big careers] Individuals may make the journey from PhD to chair on a single project.
\end{description}
There is a discussion of the features of \q{big science},
and \gls{LIGO}'s progress towards that style of working, in~\cite{collins03}, with an
extended history of the sub-discipline in~\cite{collins04}.

Because of the large costs involved and because there is usually
little immediate commercial value in this research (though of course
there are substantial long-term economic payoffs for the investing
countries), these large projects are funded at the national or
international level, so that taxpayers are the ultimate stakeholders.
Even putting aside the scientific and scholarly need
for adequate data preservation, these national investments make it
necessary for funders both to demonstrate that projects are being
efficiently exploited to produce macro-economic value, and to make the
\index{data products}\index{open data}
data products available for public use.
We discuss open data in \prettyref{s:openness}

\subsection{Data volumes}

The most immediate problem with data curation and sharing in these
scientific areas\dash though in the end not the most significant one\dash is
the data volumes involved.\index{data!volume}
The current volume of LIGO data is of the
order of hundreds of terabytes, and the data rates is expected to
grow, over the course of the project, from its current
\SI{100}{\tera\byte\per\yr} to around \SI{1}{\PBY}
(see \prettyref{tab:LIGOdatasizes}, which shows the variation in data
size for science runs three to six\footnote{In the context of
  larger-scale projects, a \q{science run} is a period when the
  equipment is run more-or-less continuously, gathering scientifically
  useful data.  Between science runs, the experiment will either be
  down for maintenance, or on a planned \q{engineering run}; data from
  engineering runs is generally stored, but is not expected to be
  useful to scientists.}).

\begin{table}
  \begin{center}
  \sisetup{table-format=3.2,table-number-alignment=center}
  \begin{tabular}{c|SSSS}
    &
    {\textbf{S3}} &
    {\textbf{S4}} &
    {\textbf{S5}} &
    {\textbf{S6}}\\
  \hline 
    L0 & 57   & 32 & 816    & 261 \\
    L1 & 8.24& 4.04  & 119 & 76  \\
    L2 & 1.55     &          &   & \\
    L3 & 0.97& 0.86  & 9.70  & 3\\
  \hline
   (duration/day) & 70 & 29 & 695 & 482 \\
  \end{tabular}
  \caption{\label{tab:LIGOdatasizes}LIGO data set size estimates in TB, and run lengths in days,
    for science runs three to six (\q{S$n$}), and various data types
    (size data taken from~\cite{LIGOdatasizeWP}; there were a total of six science
    runs in LIGO; L0 is the run's raw dataset, and L1 to L3 are progressively reduced).}
 \end{center}
\end{table}

\gls{LIGO} is just one of several other existing or planned big physics
projects, including the \gls{LHC}, the \gls{SKA}, and various \gls{ESA}/\gls{NASA}
space missions.  In comparison with these projects, LIGO's
data handling requirements are relatively modest.  The LHC 
will have data volumes of tens of
\si{\PBY}\footnote{ATLAS, one of the two larger LHC
  detectors, stores \SI{3}{\PBY} by itself; see
  \url{http://atlas.ch/pdf/atlas_factsheet_4.pdf} for some
  entertaining numbers.}
Further in the future, the SKA (which is due to be commissioned around 2020)
has predicted requirements up to \SI{1}{\tera\bit\per\second} locally and \SI{100}{\giga\bit\per\second}
intercontinentally; this involves transporting, though not necessarily storing, around \SI{1}{\tera\byte\per\minute}  or
\SI{0.5}{\exa\byte\per\yr}~\cite{taylor07}.  This is 0.05\% of the
predicted \SI{1}{\zetta\byte\per\yr} \emph{total} worldwide IP traffic for
2015~\cite{cisco11}.\footnote{%
  \SI1{\peta\byte} is \SI{1000}{\tera\byte};
  \SI1{\exa\byte} is \SI{1000}{\peta\byte};
  \SI1{\zetta\byte} is \SI{1000}{\exa\byte}; note that the unit~B
  refers to bytes, not bits}

Large-scale physical science experiments have long produced
significant data volumes, but in recent years datasets appear to be
increasing in volume and in complexity at an overwhelming rate, and
this may present a qualitatively different data management problem.
This is sometimes described in rather apocalyptic terms\dash as a \q{data
  deluge} or the like\dash and some of the challenges and opportunities
are described in~\cite{hleg10}.

\subsection{Data management styles in the physical sciences}
\label{s:styles}

\index{big science|(}
It seems useful to discuss, here, some of the distinctive features of
data collection and management in the experimental physical sciences,
since these have an impact on both the expectations for, and the
problems with, the data.

Big-science research projects have a number of relevant common features:
\begin{description}
\item[Large data sets] Such projects' data sets are \q{large} in the objective
  sense that the projects are typically so greedy for data storage,
  that their holdings are near the edge of what it is technically
  feasible to store and transport.
\item[Innovative data management] As part of the response to their
  need for large data volumes,\index{data!volume} big-science projects are often
  extremely innovative in their solutions to data management problems,
  to the extent that they are willing to work with experimental
  filesystem types, or adapt and extend operating system software or
  network transport protocols (see \url{http://lcg.web.cern.ch/} to
  get an impression of the scope of development efforts here).
\item[Specialised software] Because the instruments and their data
  sets are so complicated, these projects typically generate large
  custom data analysis software suites.  These may require
  specialised and unwritten knowledge to use, and therefore appear to
  represent a significant software preservation challenge.
\end{description}

Beyond the substantial software engineering challenges described
above, the physical sciences tend to have few \q{IT} problems, since the
communities contain plenty of people with sufficient technological nous to
address essentially all day-to-day computing-related problems,
and these communities are therefore generally reasonably
well-organised with regard to backups, storage, and basic file sharing
(see also the discussion of technological readiness in
\prettyref{s:aida}).  At the same time, however, the communities are
rather conservative from the point of view of a computer scientist,
and sometimes rather informal from the point of view of a software
engineer.  That is, the attitude to custom computing solutions is very
similar to the attitude to custom lab hardware: it may need to be
creative and experimental, but never for its own sake; it must be
stable, but is never frozen; it is accurately made, but rarely
polished.  The analogy with lab hardware and software holds to the
extent that, in the LHC community, data management groups are regarded
as detector subsystem groups; that is, they have the same general
status as the magnet or accelerator engineers, and expected to produce
agile and innovative computing services very different from the more routine,
lab-wide, provision of CERN IT services.\footnote{One of the DCC
  researchers, commenting on this report, quoted a GridPP survey
  respondent commenting that the LHC computing task: \qq{Providing resilient
    services that maintain access to data for the experiment users
    24/7\dash services are complex, bleeding edge, and are constantly
    being updated. Controlling that process, whilst also maintaining
    service up-time is very challenging}.}

The result of this is that lab software represents functional
solutions to im\-mediate-term problems, generally with flexibility
enough to respond to medium-term problems, but without much attention
being given to the imponderables of the long term, after the
experiment has completed.  It is precisely these \gls{Long Term}
preservation questions, in the OAIS\index{OAIS} sense of more than one
technology generation, that are the concern of this report.


There is plenty of prior art in this area.  See reference~\cite{ball10} for
a review of data management practice in a variety of scholarly areas,
which additionally covers several proposed life-cycle models, and
analysis techniques.
There is a similar overview in the PARSE.Insight case-studies
report~\cite{parse-33-10}, which examines data
management practice in HEP, earth observation, and social science and
humanities.  These case-studies were conducted via interviews, and
participation in ongoing efforts within the communities.  The same
project produced a gap analysis and roadmap, which make valuable reading.

This is a good place to stress that \q{big science} generally handles
its data well, and can even be regarded as exemplary (compare
\prettyref{s:aida}).  There are a few features which naturally
encourage good data management practice in the large-scale physical
sciences.
\label{s:big-science-easy}
\begin{itemize}
\item These are often relatively well-resourced projects, with plenty
  of computing experience and lots of engineering management.  There
  is lots of obvious infrastructure in the development of a large
  collaborative experiment, which gives data management an obvious
  budgetary home, where it is not competing with funding which
  directly supports researchers.
\item Astronomy and HEP projects have always produced \q{large} data
  volumes: this makes \adhoc\ data management manifestly unattractive,
  and encourages explicit data management planning and discipline.
\item The scale of these experiments means that they tend to be shared
  facilities providing documented services to their users, so that
  documented interfaces and SLAs are natural.
\item These projects rarely if ever produce commercially sensitive
  data, so that the confidentiality concerns are well circumscribed,
  concerning professional priority rather than IPR or other financial worries.
\end{itemize}
Although these features are to a greater or lesser extent specific to
this type of science, they have given rise to the notions of
\emph{data products} and explicit \emph{proprietary periods}, which we
believe would be useful in other areas, and which we discuss in
\prettyref{s:reification}.

Although it is \gls{GW} data which is our nominal focus in this report, it is
convenient to first describe general astronomy data, then distinguish
that from \gls{HEP} data\index{HEP data}, which has a somewhat different data culture,
and then describe how the GW community, which is in many ways
intermediate between the two, handles its data.

\subsection{Astronomy data}
\label{s:astrodata}

\index{astronomy data|(}
Astronomy (excluding GW astronomy for the moment) has probably the
most straightforward data management practices in the physical
sciences.  When an optical telescope takes an image (or a spectrum,
which for our present purposes is technically equivalent to an image), either as part of
a systematic survey of the whole sky or as a pointed observation
requested by an astronomer, the image is typically moved from the
telescope's detector straight into its archive, from where it can be
later retrieved by the astronomer, accompanied by automatic or
manually-added metadata.

Non-optical astronomers (covering the rest of the spectrum,
from radio to gamma rays), and most satellite missions, have a
somewhat more complicated route from observation to image, and a
broader set of data products, but have essentially the same model, and
the same discipline and expectations around archives.  From the point
of view of data management therefore, we can elide the differences
between the various branches of astronomy.
Gravitational wave and neutrino astronomy, in contrast, are not
studying the electromagnetic spectrum, and partly as a consequence
their study more closely resembles particle physics (see
Sects.~\ref{s:hepdata} and~\ref{s:gwphysics} below).

Most large telescopes, satellites and instruments\footnote{An
  \q{instrument} in this context is the light-sensitive detector
  attached to the telescope or satellite optics (the camera, in
  effect).  It is replaceable or swappable, and regarded as a separate
  piece of engineering from the telescope.  The days when observers
  would travel to the telescope carrying their own instrument are now
  largely past.} operate partly or exclusively
according to a model in which astronomers are awarded \q{telescope
  time}, ranging from a few hours to a few nights, as the results of
competitive bids closely analogous to grant bids.  The resulting data
generally has a proprietary period\index{proprietary data}, extending
for perhaps 12, 18 or 24 months after the data is taken, during which
only the observer who requested it can retrieve it, but after which it
automatically becomes retrievable by anyone (\q{embargo} would be a
better term, though unconventional).  Similarly, instruments
built by consortia generally have proprietary periods during which the data is
only available to consortium members.  The proprietary periods are
partly for the benefit of the consortium individuals\dash it is their
reward for the initiative and possibly decadal effort of building the
instrument\dash but they are also a pragmatic reflection of the length of
time it may take to calibrate and validate acquired data, ready for
deposit in an open archive.  As a result, the lengths and terms of
proprietary periods are the subject of negotiations between the
instrument builders and their ultimate funders, though the
negotiations are always about the length of the delay before a general
data release, and never question the necessity for the release itself.

\gls{NASA} missions now typically have 12-month proprietary
periods, but this has varied historically, and for example the 1990
COBE mission, which included significant technological novelty, and
whose performance was therefore rather unpredictable, had a 36-month
proprietary period.

Not all instruments have formal release plans, and the proprietary
periods that exist may be adjusted informally.  Caltech\index{Caltech} is one of the
few private institutions which is rich enough to own, or have a
significant share in, world-class telescopes (Palomar and
Keck).\index{private facilities}
It has no declared policy on data management or data sharing, beyond a
broad tacit expectation that data will be published as appropriate for normal
scientific practice.  As a second example, during the \q{science demonstration phase}
of the commissioning of the Herschel\index{Herschel} telescope (that is, the
last commissioning phase, verifying that the science goals were
achievable), the instrument team invited\footnote{See \q{Herschel
    Observers' Manual} \S1.1.4,
  \url{http://herschel.esac.esa.int/Docs/Herschel/html/ch01.html};
  thanks to Haley Gomez for bringing this to our
  attention.}\label{s:herschel}
 observers to nominate part of
their scheduled observations to be performed early, during this still-experimental
commissioning phase.  When the observations proved successful
(as they generally were), the observing teams were given the choice either of making
the data immediately public, in time for the opening of the
Herschel archive and a journal special issue, and having the
observation time re-credited to them; or else retaining the 12 month
proprietary period\index{proprietary data} without the re-credit.

Image data is the archetypal astronomy data, and is generally stored as files, but
another important category is the astronomical \q{\gls{catalogue}}, of
object positions, spectra and other properties, usually stored in relational
databases.  Astronomical archives range from quite small ones (at one
extreme, a small specialised instrument may have its \q{archive} consisting of
a file server looked after by a graduate student) to very large
professionally managed archives which are both the primary sources of
some data sets, and mirrors of others.

Astronomy data is potentially very long-lived.  Although astronomers
are naturally drawn to the newest instruments with the greatest
sensitivity, it is not unusual to draw on relatively old archive data.
In most cases, this will be still be born-digital data, but
digitised versions of century-old astronomical plates are used in
precise astrometry, and to identify the precursors of supernovae and
other one-off events (see for example the Edinburgh
SSA\footnote{\url{http://surveys.roe.ac.uk/ssa/}}, further discussed,
with background, at~\cite{hambly01}; and a more discursive account of
plate scanning, including discussion of some of the archival
challenges, in~\cite{jones01,bhattacharjee09}). 
Even babylonian and ancient chinese astronomical data\index{Babylon} has
been used for contemporary science, helping measure the rate at
which the earth's spin rate, and thus the length of day, is
changing~\cite{stephenson95}; similarly, 3\,000-year-old egyptian data
has been used to measure the change in the orbital behaviour of the three
stars in the Algol system~\cite{jetsu12}.  The cosmos
changes slowly on our timescales, so that the great majority of
astronomical observations are repeatable; the exceptions are those
cases where long time-bases are necessary (precise astrometry) or where
the object of study is a one-off, and therefore unrepeatable, event
such as a recent or historical supernova.

Astronomy data is
also intelligible in the long term: although
untranscribed babylonian tablets can only be read by specialists, contemporary
astronomers can basically understand the data published in Kepler's 1627
\emph{Rudolphine Tables}, and with some assistance can understand the
content of the 11th- to 12th-century Toledan Tables~\cite{toomer68}.
Although biologists might be able to make
similar claims with respect to, for example,
\ifxetex Linnæus's\else Linn\ae us's\fi\space observations,
it is hard to find equally long-lived data in the physical sciences,
or born-analogue physics data where there is a similar contemporary pressure
for digitisation.

There is essentially no file-format problem in (electromagnetic) astronomy,
since the \gls{FITS} format is universal~\cite{fits3,pence10}.  Though not
perfect, this is a relatively simple and well-defined format,
combining binary or table data with keyword-value metadata.

\label{s:rawdata}
Astronomical data also has a well-developed notion of \emph{\gls{data
  products}}\index{data products}.  These are datasets which
contain, not \gls{raw data}, but data which has been processed to a greater
or lesser extent.
We can distinguish at least three levels of data in this context; most
large instruments will have more than one level of derived products.
\begin{description}
\item[Raw data]
This is the lowest-level data, consisting of the direct output of a
detector or other instrument, or the raw satellite telemetry.
This data is made meaningful only by processing with
software which is to some extent specific to the equipment in
question.  Though it will be preserved as a matter of course, it is
rarely published, nor used by, nor useful to, other 
researchers, except in unusual circumstances.  In the case of a
particularly subtle effect\dash or less commonly, a debate over a
theoretical analysis or calibration\dash a researcher might return to the \gls{raw data},
but this will generally be done with the collaboration of the
instrument scientists, and may be otherwise infeasible, to the extent that any
results obtained without such insider knowledge might not be
believable by the broader community.
\index{raw data}
\item[Data products]
After it is gathered, (raw) data must be processed (\q{reduced}) to
turn it into scientifically meaningful numbers (interpreting
engineering or telemetry data streams, and calibration) and to
remove various instrumental and observational artefacts.\index{data products}
Data products are usually 
made available in standard formats (in astronomy, generally FITS
files), whereas raw data, if it is made available at all, may well be
in an instrument-specific form.
\item[Publications]
Sitting above the data products is a class of high-level 
outputs, including scientific papers, and other peer-reviewed outputs
such as published \gls{catalogue}s.  Journal articles are curated at
publisher sites and the \gls{ADS}, and article preprints at the
\gls{arXiv} (cf \prettyref{s:biblio}).  Modest volumes of data can be published as digital
appendices to journal articles in, for example, Astronomy and
Astrophysics Supplements; these are curated at the journal and at \gls{VizieR}.
\end{description}

It is the data products\index{data products} which are the outputs
which are sufficiently free from observational artefacts to be the
starting point for scientific analysis (high-level products are
sometimes referred to, informally, as \q{science data}), and which
represent the class of data which is naturally archived, most
carefully documented, and which will eventually be made public.  There
may be multiple levels of data products, with lower-level products
carrying more information, but using which requires more detailed
knowledge of the subtleties of the instrument and its processing \gls{pipeline}.  To a
much greater extent than is true for HEP data, for example, the
highest level astronomy data products are both useful and generally intelligible\dash
everyone is, after all, looking at the same sky\dash but researchers
will often use intermediate-level products, if they can invest the time to
learn about them, or have collaborators who have experience with them.
Those researchers who are more intimately involved with an instrument will
be comfortable using lower-level data products, because they will have
the knowledge which enables them to run, or experimentally re-run, the pipelines in a
scientifically meaningful way.
That said, in OAIS\index{OAIS} terms, astronomical data can be
characterised as having a broad \gls{Designated Community} and well
understood \gls{Representation Information}.

Publications are in the province of libraries and similar
repositories, and are not considered further in this report.

Optical astronomy (that is, with observations made using visible
light) has the most straightforward data, so that the distinction
between \gls{raw data} and data products is slight to the point of being
rather artificial: astronomers reusing optical data would expect to
recalibrate the raw or nearly raw data, and would not anticipate having difficulty
doing so.

We conclude with some examples: A typical telescope archive is the UKIRT archive at
\url{http://archive.ast.cam.ac.uk/ukirt_arch/}; there are several
image and spectrum archives at the Royal Observatory Edinburgh's Wide
Field Archive Unit\footnote{\url{http://www.roe.ac.uk/ifa/wfau/}}; and there is a large
collection of \gls{catalogue}s available at \gls{CDS} (see
\prettyref{s:cds} below).

The \gls{ESA} Hipparcos astrometry
mission\footnote{\url{http://www.esa.int/science/hipparcos}}\index{Hipparcos}
flew between 1989 and 1993, and produced a high-precision catalogue of
100\,000 stars~\cite{hipparcos97}.  The catalogue is available online
as queriable databases at
ESA\footnote{\url{http://www.rssd.esa.int/index.php?project=HIPPARCOS}}
and \gls{CDS}, as CDs, and as PDFs which match the catalogue's
17-volume printed version. The printed version is an interesting case:
as discussed in the catalogue (vol.\,1, \S2.11.3), the printed
pages are designed with a per-page checksum, to help with re-scanning
the catalogue from paper, in the presumed-likely future case that the
digital version becomes unreadable and only copies of the paper book
survive.  The Tycho catalogue, from the same mission, comprises around
20 times the number of stars, at lower precision, and is only available
online.\index{astronomy data|)} 

There is some discussion of preservation costs in \prettyref{s:preservation-costs}.

\subsubsection{Strasbourg Data Centre (CDS) as a disciplinary repository}
\label{s:cds}

\gls{CDS} is a large disciplinary repository for astronomy~\cite{genova00}.
It stores a broad range of catalogues, of various sizes, in its \gls{VizieR}
service (see \cite{ochsenbein00} and
\url{http://vizier.u-strasbg.fr/}) and provides a large
librarian-curated collection of data from, measurements of, and
references to, individual astronomical objects.  It cooperates closely with \gls{ADS}.

CDS was created, and is supported, by the french agency in charge of
ground-based astronomy\dash first CNRS/INAG then CNRS/INSU\dash as a joint
venture with Strasbourg University.  The main support is through
permanent positions from the CNRS/INSU and the University
(researchers, computer engineers, and specialised librarians), with
additional contracts supported by funding from various sources.

CDS is administratively located within a research structure,
Strasbourg Observatory, providing an active research environment for
CDS astronomers.  The preservation aspects have never been separated
from the provision of services and the maintenance of local expertise
on data management and preservation.\footnote{We are most grateful to
  Fran\ccedilla oise Genova, of CDS, for this discussion of CDS's history and support.}

This can be seen as an example of a very successful disciplinary
repository.  There appear to be several key features of this success.
\begin{itemize}
\item CDS has established, and actively maintains, international leadership in
  the curation of astronomical data, by virtue of collaborating widely
  and investing effort in projects (such as the \gls{IVOA}) which
  support and promote data sharing.
\item As a result of the intimate relationship between the repository,
  the observatory and the university (to the extent that the
  boundaries between the three can seem rather vague to outsiders),
  CDS personnel have practical knowledge of how their data is used,
  and what researchers need.
\item The core funding for CDS comes from the french state, but it is
  conceived as an internationally visible project.
\end{itemize}

\subsubsection{Collaborations in astronomy}

The most visible collaborations in astronomy are large terrestrial and
satellite-borne telescopes and other instruments.  At the risk of
oversimplifying, these are generally not the many-person
collaborations usual in \gls{GW} or \gls{HEP} physics, but are instead
facilities created by space agencies or consortia of national funders.
Although they are highly innovative leading-edge facilities, they are
not seen as \emph{experiments} in the same way as \gls{LIGO} or the
\gls{LHC} are (massive) items of specialised hardware built to answer
a delimited set of scientific questions.  They are instead
\emph{observatories}: the data management in these facilities is part
of their general operating infrastructure, and the research and
research data they produce is \q{owned} (at least in an academic
rather than a legal sense) by the scientist \emph{users} of the
facilities, rather than the facility itself.

Astronomy does however have a variety of data-analysis
collaborations.  These are semi-formal collaborations concerned,
mostly, with multi-wavelength studies of multiple archives, and
include for example UKIDSS-UDS,\index{UKIDSS} the Herschel\index{Herschel} Atlas collaboration,
HerMES\index{HerMES} and
GAMA\index{GAMA}\footnote{\url{http://www.nottingham.ac.uk/astronomy/UDS/},
  \url{http://www.h-atlas.org/}, \url{http://hermes.sussex.ac.uk/},
  \url{http://www.gama-survey.org/}}.
These have between 20 and 60 collaborating members scattered over
perhaps a dozen institutions but, crucially, no \q{corporate}
existence, and little or no direct funding.  Instead, they are funded
indirectly via individual fellowships or rolling grants: participation
in the collaboration might be a strong feature of a grant application,
but it is not the collaboration as such that receives the direct
support.  They do have governance structures, but these tend not
to be particularly formal, because they remain small enough that there
is little perceived need.
These collaborations exist to derive high-value derived data products
from the lower-level data products of the archives they are analysing
(for example Herschel is an \gls{ESA} observatory mission: this means that
individuals can bid for observations, but that ESA does not have it as
part of its remit to provide more than minimally reduced science
data).

The collaborations distribute their results in papers, and associated
datasets; they typically build archives to support and distribute
their work, but there's no expectation (beyond the usual cooperative
academic norms) that they will help others work on the data, or
release it.  It is hard to see how there could be such an expectation,
much less an obligation, since they receive little direct funding, and
their indirect funding comes from a multinational set of entities with
potentially very different \gls{DMP} policies.

\subsection{High Energy Physics data}
\label{s:hepdata}

\index{HEP data|(}Astronomy is essentially an observational science: telescopes,
their optics, and the detectors which hang off them, are constructed
to create a path from nature to data which is as nearly as possible
unmediated.  This means that it is both reasonably obvious what things
are to be archived, and that the nature and processing of
observational artefacts are well and commonly understood.  This means
that astronomy, unusual in the physical sciences for needing to preserve
data long-term, is in the happy position of having its data readily preservable.

\gls{HEP} data is different.  HEP is a participative science, where
objects ranging in size from electrons all the way up to nuclei are
disassembled, and data about the messy results of this
disassembly is examined to retrieve information about the interior structure of the
original.  This reconstruction from collision data depends on a
shifting engineering understanding of rivers of data,
out of instruments which are one-off works of art, designed and
assembled by a thousand-strong community, close-packed
into a detector the size of a small cathedral, attached to a machine
with its own postcode.

The result of this is that HEP data analysis is rather tricky, with
many steps between data and science, each of which depends on software
which encodes a detailed understanding of the data's provenance.  In
consequence, although HEP data is typically distributed with multiple
levels of reduction, almost none of these levels (with the exception
of formal publications) are straightforwardly suitable for long-term preservation.  This
is because interpretation of this data is heavily dependent on
software, the use of which requires detailed experimental knowledge
which it may be infeasible to preserve.  In OAIS\index{OAIS} terms, the designated
community is tiny because the \gls{Representation Information} is
hugely complex.

In addition to this, HEP data has a considerably shorter shelf-life
than astronomy data, as discussed above.  In contrast, old HEP data is
typically made redundant by new data, obtained from more powerful accelerators.
Also in contrast to astronomy data, HEP data is not expected to be
generally intelligible for very long: two- or three-decade old data
might potentially be useful or intelligible, but much beyond that
would count as archaeology.  At the risk of being whimsical, we can
compare the roughly millennial lifespan of astronomical data with the
roughly three-decade lifespan of HEP data, and conclude that the latter goes
\q{off} about 30 times faster than the former.   Although facilities make very
considerable efforts to manage data safely while an experiment is running,
there is little real pressure to preserve HEP data into the long
term.

Of course, things are not quite as straightforward as that in fact.
(i)~The \gls{LHC} gains interaction energy at the expense of a messier
collision, so there are potentially some features that will be
detectable in one dataset (for example the \gls{HERA} p-e data) which
would not be findable in the LHC.  While interaction energy is the
most prominent metric of an accelerator's performance, it is not the
only one, so that larger accelerators will not render smaller ones
obsolete as inevitably as we may have suggested above.
Similarly to this, (ii)~data reduction errors may be
dominated by theoretical uncertainties rather than experimental ones,
and these will only be improved, and the data re-reduced, after the
experiment is over.  Finally (iii)~there are no accelerators bigger
than the LHC currently scheduled, so that this dataset may remain the
highest-energy one for a relatively long time.  The archaeology is
illustrated in~\cite{curry11} and the problem further
explored in~\cite{south11}, which also discusses the HEP community's
developing plans for data preservation.  Qualifications
notwithstanding, the overall timescales in HEP are shorter than in
astronomy, and the solutions described in~\cite{south11} are concerned
with prolonging a continuous low-level relationship with a dataset
rather than being able to return to a dataset cold.

Unlike astronomy, HEP has for the last few decades been organised into
larger and larger collaborations, and these collaborations have
developed intricate, and socially fascinating, cultures for managing
this.  The two larger instruments at the LHC, \gls{ATLAS} and
\gls{CMS}, each have author lists of order 3\,000 people, so that the
various CERN collaborations account for around 10\,000 research-active
individuals.  There is extensive discussion of the history and
structure of the LHC collaborations in~\cite{boisot11} and in the
outputs of the \textsc{pegasus}
project,\footnote{\url{http://www.pegasus.lse.ac.uk/research.htm} and
  in particular \cite{kyriakidou09}} but many of
the collaborations' relevant organisational features are echoed in the \gls{GW}
community: this is discussed in \prettyref{s:consortia} and we do not
discuss them here.\index{HEP data|)}




\subsection{Gravitational wave physics}
\label{s:gwphysics}

\index{data!gravitational wave|(}The gravitational wave community has astronomical goals, but in the
scale of the \gls{LIGO} project, and in the amount of novel technology involved,
as well as in the fact that many of the personnel involved came
originally from a HEP background, the project's culture more closely
resembles that of a HEP experiment than of an astronomical telescope.
We discuss some specific features of LIGO data in~\cite{carozzi10};
here we discuss where \gls{GW} data, and the discipline's organisation
structure, fits on the spectrum between astronomical and HEP data.

\subsubsection{Gravitational wave consortia}
There are three principal sources of recent GW data available to UK researchers: LIGO, GEO600 and
Virgo. There are other detectors which are either smaller efforts (in
terms of consortium sizes), which have stopped taking data (TAMA-300),
or which are still at the planning stage.  See~\cite{pitkin11} for an
overview of current detectors, and of detector physics.

\label{s:consortia}
\gls{LIGO Lab} is a collaboration between Caltech and MIT,
which designs and runs three interferometers in Hanford, WA,
and Livingston, LA, in the US. \gls{GEO} is a German/British
collaboration, which runs the \gls{GEO600} interferometer.
The three LIGO interferometers were shut down in October 2010 to refit
for \gls{aLIGO};
the GEO600 interferometer is still currently running.
The \gls{LSC} is the result of a network of \glslink{MOU}{Memoranda of
  Understanding} between LIGO Lab (or more loosely the LSC) and
multiple other institutions of various size.\footnote{The MOU which
  created the LVC is at~\cite{lsc-m060038}, but MOUs are not
  routinely made public.}  These relationships
involve hardware, resources, and data access of various types.  Most
typically, the resources in question are personnel, and an institution
such as a university physics department, which wishes access to LIGO
data, will contribute in return fractions of staff from permanent
staff, through post-docs, to PhD students, for a broad spectrum of
activities including  data analysis, instrument fabrication and
shift-work in the detector control room.  However
in some cases, the MOUs are concerned with data swaps, and set up
limited data releases with other scientists: for example, there are a
few MOUs between the LSC, Virgo and other observatories, which
describe what data is to be shared, in what volumes, and the outline
authorship arrangements for any subsequent papers.  GEO's MOU
describes a particularly close relationship with LIGO Lab, but most of
the MOUs are broadly similar to each other, and the process of creating one is by
now streamlined.  In total (as of June 2010),\footnote{The definition
  of LSC membership is included in~\cite{lsc-m050172} and the
  construction of the author list in~\cite{lsc-t010168}.}
the LSC consists of a little over 1300
\q{members};
of these, 615 spend more than 50\% of their time dedicated
to the project and so have a place on the LSC author list.

The term \q{LIGO} has a number of not quite equivalent meanings:
sometimes it refers to LIGO Lab, sometimes to LIGO Lab plus the LSC,
and the phrase \q{LIGO detectors} is generally understood to refer to
the LIGO Lab and GEO detectors.

The Italian/French \gls{Virgo} consortium has its own detector and
analysis \gls{pipeline}, and has a data-sharing agreement with the
LSC, represented by the \gls{LVC}.\footnote{The term \q{LVC} is not an
  initialism.  It colloquially refers to the data-sharing
  agreement~\cite{lsc-m060038} and joint meetings between the LSC and
  the Virgo Collaboration.  Though there are \q{LSC/Virgo
    collaboration groups}, there is no formal big-C Collaboration.}
As with LIGO, the Virgo detector will shut down between 2011 and roughly 2015.
Virgo has 246 members (with a slightly different definition from the LSC), and GEO600 around 100.

\begin{figure}
\begin{centering}
\includegraphics{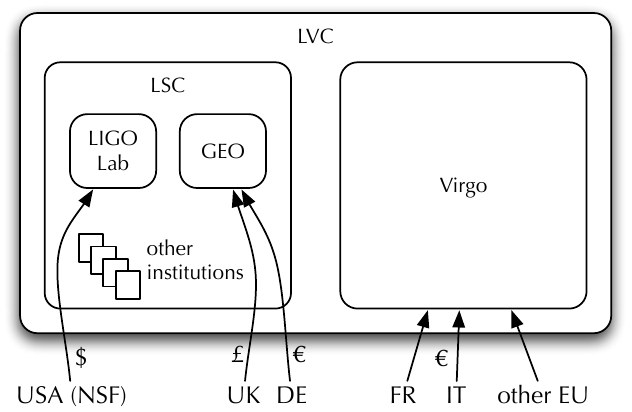}\\
\end{centering}
\caption{\label{f:lvc-diagram}The relationships between various GW consortia.}
\end{figure}

There is an attempt to summarise these relationships in
\prettyref{f:lvc-diagram}.

These experiments have a common purpose:
they exist to detect signatures of gravitational waves, which are confidently predicted by the General
Theory of Relativity, but the actual observation of which would be a
major scientific event (there exists an LSC data processing flowchart
which includes the not entirely serious branch \qq{Call Stockholm!}).

Gravitational waves are sufficiently weak, however, that the existing
equipment will not become sensitive enough to have a good chance of detecting them until after
its refit, which began in late 2010 (when the project entered the phase known as
\gls{aLIGO}), and which is scheduled to be
completed when the new detectors are commissioned in 2015.

\subsubsection{GW data}
\label{s:gwdata}

Although the consortia have (as expected) announced no detection
so far, they nonetheless produce a large volume of auxiliary data,
representing background and calibration signals of various types, and
this, together with the core data, means that the LSC collectively
produces data at a rate of approximately one \si{\PBY}.

We can readily identify the levels of data which were discussed in \prettyref{s:astrodata}:
\begin{description}
\item[Raw data]
The lowest-level GW data consists of the signals from the core
detectors.\index{raw data}  This data is made meaningful only by processing with
software which is completely specific to the detectors in question.
This is stored in \q{frame format}, which is a very simple format
intelligible to all the primary data analysis software in the
community, and which is multiply replicated across North America,
Europe and Australia.  Although the disk format is common, the
semantic content of the raw data is
specific to detectors and software, so that preserving it long-term would
represent a significant curation challenge.
\item[Data products]
The raw data is processed into calibrated \q{\gls{strain data}}, which is the
data channel in which a GW signal will eventually be found (this is
possibly, but not necessarily, also held in frame format).  This is
the class of data products\index{data products} 
which will eventually be made public.
Unusually, it turns out that
GW raw data is in a semi-standard format, and the data products are
specific to the analysis \gls{pipeline} which produced them.
\item[Publications]
Sitting above the data products is a class of high-level data
products, scientific papers, and other peer-reviewed outputs.  The GW
projects have announced no detections of gravitational waves, but have
nonetheless produced a broad range of astrophysically significant
negative results~\cite[\S6.2]{pitkin11}.
\end{description}

As with the general astronomy data products discussed in
\prettyref{s:astrodata}, the distinction between the \q{raw
  data}\index{raw data} and
the \q{data products}\index{data products} is that
the latter datasets, alongside their supporting documentation, will be
available for use and reuse by scientists who do not have an
intimate connection with, and knowledge of, the instrument.

Both the \q{data product} and \q{publication} groups are broad classes of
objects.  The practical boundary between them is clear, however: what
we are calling \q{publications} are entities such as journal articles or
derived catalogues whose long-term curation is not the responsibility
of the LSC data archive, though they may be held in some separate LSC
paper archive, which is as such out of scope for this project.

\subsubsection{Gravitational wave data releases}

Because the LSC has not announced the detection of any signal so far, and because the
data will remain proprietary to the consortium until well after such
an announcement\index{proprietary data}, there are no distributed data products
so far, and so the issues surrounding formats and documentation have
not yet been addressed.  However it is the eventual public data
products which are the highest-value outputs from the experiment, and
which are the products which it will be most important to archive
indefinitely.

At present, LIGO data is available only to members of the \gls{LSC}.
This is an open collaboration, and research groups which join the LSC
have access to all of the \gls{LIGO}
data\footnote{\url{http://www.ligo.org/about/join.php}}.  In return,
they contribute personnel to the project (including for
example people to do shift-work manning the detectors), and accept the
collaboration's publication policies, which require that all
publications based on LIGO data are reviewed by the entire
collaboration, and carry the complete 800-person author list.  At
present, and in the future, data which is referred to by an LSC
publication is made publicly available.  See
\prettyref{s:stuart-and-roy} for further details on LIGO's \gls{DMP} plan.
\index{data!gravitational wave|)}

\subsubsection{Summary: big-science preservation challenges}

In the three sections above, we have tried to describe both
differences and commonalities between three large-scale scientific
disciplines.  Possibly the big\-gest difference between the three areas
is that high-level astronomical data products are much more generally
intelligible than even the highest-level HEP products.  In each case,
however, we have a ladder of reasonably well-defined data products,
with each rung generated from the lower ones by sophisticated data
reduction pipelines.

The situation is not as rosy, from the point of
view of long-term preservation, as this account may suggest.  Because
the pipelines have developed organically over a number of years, under
the influence of experience with earlier versions and increased
understanding of the instrument, the knowledge they represent is
sometimes encoded within them in a less structured way than would be
desirable.  Sometimes, metadata is encoded in filenames, or in
configuration files, or wikis, or even private emails.
Of course, one could simply argue that this information should be
documented better, but it would be hard to argue that the costs of
this work would be justifiable, to service a future theoretical need that few
believe would even become an actual one.
In consequence, although the resulting data product will be regarded as
perfectly reliable, it may be infeasible to redo the analysis other
than by preserving and rerunning the pipeline software (even if it
were feasible, it would be prohibitively expensive, and rarely seen as
valuable; see also \prettyref{s:preserve-raw}).  For this reason, software preservation\index{software
  preservation} has some role in the overall data preservation
strategy.  However it is not clear to us what this role should be, and
the thorny issue of software preservation\index{software
  preservation} is addressed at greater length in \prettyref{s:sw-preservation}.
\index{big science|)}

\subsection{A contrast: social science data}
\label{s:socscidata}

It is possibly instructive to contrast the data management practices
discussed here, with the very different problems faced by data
managers in the social sciences. In~\cite{van-den-eynden10},
the authors survey a number of social science projects\index{social sciences}, with a particular
focus on two large (for the social sciences) programmes funded by the \gls{ESRC}
(the UK social science research council) with substantial
responsibilities for data preservation and sharing.\footnote{This work was part of the \q{Data
  Management Planning for ESRC Research Data-Rich Investments} project
  (DMP-ESRC)
  (\url{http://www.data-archive.ac.uk/create-manage/projects/JISC-DMP}),
  funded by JISC, like the present project, as part of the Managing
  Research Data programme.}

For the ESRC projects, the artefacts being stored are simple things,
at the level of \gls{Content Information}: they are conventional Word
documents and audio files, rather than the heavily structured and
still somewhat experimental big science data objects.  The ESRC archive
contents will remain broadly intelligible to future researchers,
without much archive-specific effort to define \gls{Representation
  Information} or a \gls{Designated Community}.  In contrast to this
simplicity, however, the ESRC archives have to cope with a broad range
of associated contextualising metadata, which is different for
different projects, and inconsistently or incompletely specified by
the originating researchers, perhaps as an afterthought.  This makes
archive ingest a complicated problem, in contrast to the big science cases,
where archive ingest fundamentally involves little more than copying a
self-contained set of artefacts from working storage to some
preservation store.  In particular, the ESRC projects have a
complicated set of anxieties about copyright, IPR, confidentiality,
anonymization and consent; while \gls{LIGO} cares intricately about data access and
security, it does so in the rather formal context of professional
ethics rather than family secrets.

This illustrates two further notable differences between physical
science data and that of social science or broader archival resources.

Firstly, the responsibility for ESRC data
in practice lies with more junior researchers, helped by part-funded
archivists~\cite[\S\S5.2.1 \& 5.4]{van-den-eynden10}.
For big science projects, it is funders and senior
collaboration members who drive the preservation efforts.

Secondly, essentially all physics data is born digital and complete,
meaning that all of the information to be archived is present at the
time of deposit.  Of course, this is not complete from the point of
view of reproducibility (that requires journal articles and personal
knowledge) and does not discount the subsequent addition of subjective
metadata as finding aids, but it is completely specified from the
point of view of conventional future analysis.  The distinction is that
experimental data is a complete and objective account of everything
that was believed to be relevant in recording a physical event which
happened at a specific time.  One can disagree with the experimenters'
beliefs about completeness (this shades into questions of
reproducibility and tacit knowledge), complain that some details might be recorded in
notebooks rather than digital records (more true of lab-scale than
facility-scale experiments), or in extreme cases argue about the
nature of objectivity, but a natural science experiment has a
much clearer boundary, in space, time and documentary extent, and so a
more natural expectation of documentary completeness, than will be
usual for an experiment in the social or human sciences.
This is different from the
traditional archive problem, where the problems of interpretation are
more visible and acknowledged, and the problem of incompleteness more evident.%
\footnote{For a vivid and illuminating discussion of the complications
and physicality of reproducing experiments, see~\cite{collins01} and
references therein (by coincidence, this describes observations
amongst gravitational wave experimenters in Glasgow); that discussion
is reprised in a larger context in \protect\cite[ch.35]{collins04}.
The question of tacit knowledge is discussed at length in~\cite{collins07}.
For a discussion of different types of reuse, see~\protect\cite[\S3]{bechhofer10}.}

The summary is not that the ESRC or the big science archives have an
easier job overall, but that the complications express themselves in
different parts of the mapping from OAIS abstractions to local fact.
Big science archives must preserve large complicated objects for a
hard-to-describe \gls{Designated Community}, but because they are
essentially always project-specific archives, their implementation does not have to be
generic, and many of the ingestion issues can be baked into the
original archive design.

\subsection{Babylonian data management (less contrast than you'd think)}
\label{s:babylon}

\newcommand\bce[1]{#1\,\textsc{bce}}
\newcommand\ce[1]{#1\,\textsc{ce}}

Contemporary astronomy began, in the west, in Mesopotamia in the fifth
and fourth \bce{centuries}.\index{Babylon}  Although earlier datasets exist\dash the
\emph{Venus tablet of Ammisaduqa} is a cluster of \bce{7th\,C} copies of
17th\,C or 16th\,C data recording the rise times of Venus over a 21 year period\dash
these earlier omen texts seem to have been preserved for largely
cultural reasons.\footnote{See \cite{aaboe91} for background and
  further references, and \cite[ch.4]{hobson09} for very detailed
  discussion of the physical tablets. The precise date of the
  observations is of considerable scholarly interest, since an agreed date would
  provide an absolute fix for the otherwise relative chronology of the Late
  Bronze Age Near East.}

\begin{figure}
\checkoddpage
\hbox to \columnwidth{%
\ifoddpage\else\hss\fi
\hbox to \overallwidth{\includegraphics[width=\overallwidth]{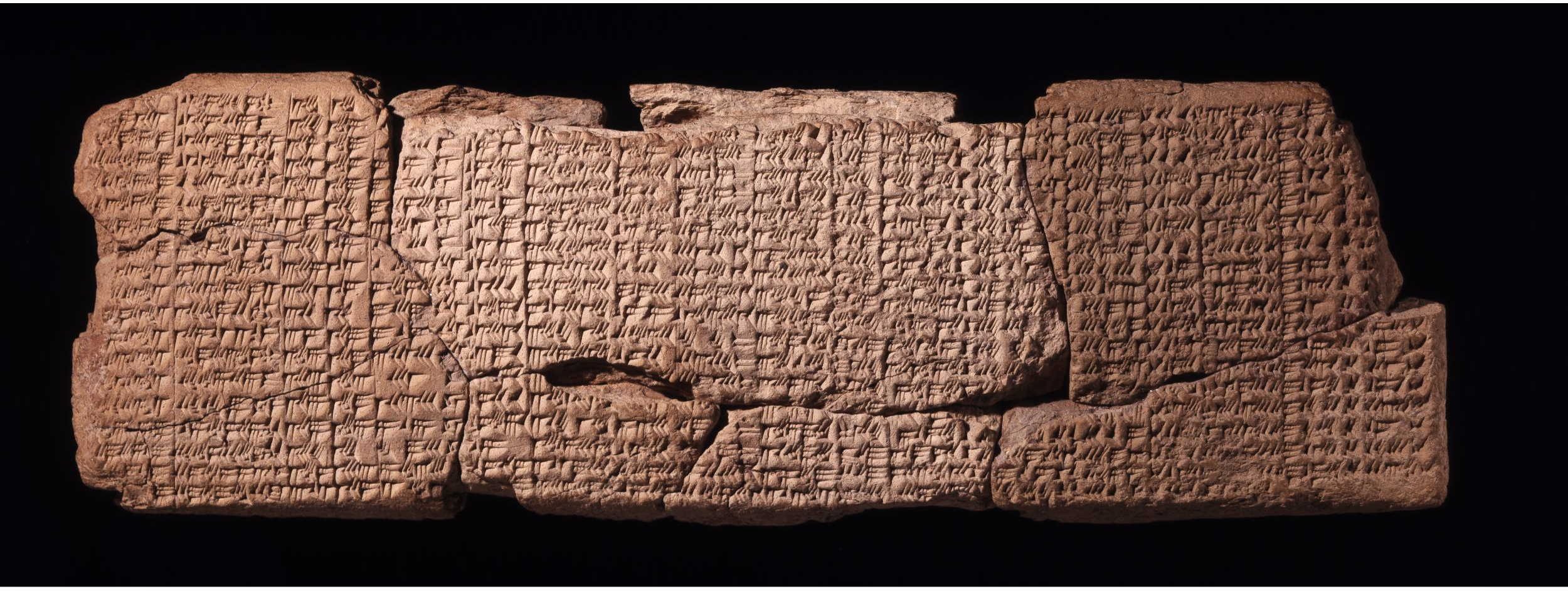}}%
\ifoddpage\hss\fi
}
\caption{\label{f:spii}Calculated ephemeris for the period \bce{104}
  March 23 to \bce{101} April 18, written on Seleucid year 209, month
  IX, day 18 (\bce{103} December 20?).  Comparison with a JPL ephemeris
  shows that the text conjunction times remain within a couple of hours of the
  correct values, with an offset attributable to an error in the
  initial value.  For detailed discussion, see \cite{aaboe74}.
  British Museum item Sp-II.52,
  \textcopyright Trustees of the British Museum.}
\end{figure}


Distinct from these, there is a large set of 4\range500 other texts,
ranging from \bce{4th\,C} to \ce{75} with a smattering going back as
far as \bce{mid-8th\,C}, and spanning the development of Babylonian
theoretical astronomy during the \bce{4th\,C}.  These are a
mixture of observations, calculated ephemerides (such as
\prettyref{f:spii}), and telegraphically obscure
technical documentation.  The observation texts\dash \q{astronomical
  diaries}, forming the majority of the texts\dash describe in sequence
celestial and meterological observations, daily commodity prices,
river levels, and topical events. The observations of the Sun, Moon
and planets were of good enough quality, and preserved over a long
enough time, that when babylonian mathematical models were fitted to
them they produced values for the synodic and anomalistic months and
(implicitly) the orbital periods of the planets, which are very
respectably close to their currently-determined values (out by a
factor of $3\times10^{-7}$, in the case of the synodic month).  These
were used to predict the first and last appearances of planets, and
the times of lunar (but not solar) eclipses.

The information in these texts is sometimes available on multiple
tablets, although it is not clear whether these duplicates
were backups, mirrors, or media refreshes.  Many tablets have
acquisition metadata, added in ink by the archives, millennia apart,
in Babylon and Bloomsbury.

It is clear that the tablets that have survived represent only a small
fraction of the total.
but both the data, and the mathematical technology that reduced the
data and generated the ephemerides, were available and fully
intelligible to Hipparchus (\emph{c}~\bce{150})\index{Hipparchus} and, either via him or
directly, to Ptolemy (\emph{c}~\ce{150}).\index{Ptolemy}  The Babylon Data Centre was
still active in the first \ce{century}, though funding cuts meant new
acquisitions were by then minimal, and it was operating in the collapsing
ruins of the desert city.

The \gls{Content Information} in the texts is sufficiently well
preserved that if the texts can be dated at all (in some cases through
contemporary ingest metadata), they can generally be dated to
the very day; the technical \gls{Representation Information}, in
contrast, is so terse as to make sense only after the procedure being
documented is reconstructed from the Content.  The cuneiform presents
a challenge, but once this has been transliterated, the datasets are
fundamentally intelligible to current astronomers.  The
preservation strategy is a daring one: by effectively founding western
astronomy, and arranging that the data was preserved just long enough that
it could be taken over by the (hellenic) successor
civilisation,\footnote{This can be classed as a \q{high-risk} data
  preservation strategy, and is not included amongst this report's
  Recommendations to STFC.}
the babylonians ensured that their coordinate system (based on the zodiac)
and number system (with angles in degrees, subdivided into base-60
fractions) would still be in use by astronomers 25 centuries later.

\subsection{Bibliographic repositories}
\label{s:biblio}

Though it is not strictly data, it seems useful to make parenthetical
mention of the big science communities' literature repositories, since
they seem to illustrate the way in which the communities have learned
to act collectively.

The preprint archive at \texttt{arXiv.org} started in 1991 as an
electronic version of the long-established practice of distributing
preprints of accepted journal articles around the high-energy physics
community, by post.  It currently receives around 6000 submissions per
month, predominantly in HEP, astronomy, condensed matter physics and
mathematics; it probably receives copies of nearly 100\% of the HEP
community's
output.\footnote{\url{http://arxiv.org/Stats/hcamonthly.html}}
Authors most typically submit papers at the point when they have been
accepted by the journal, but some submit earlier versions, and a few
are not further published at all.  Although the journals are still
providing an \emph{imprimatur}, many papers are now principally read
as preprints, and many journals permit citations by arXiv reference.
ArXiv is supported by requesting contributions from its heaviest
institutional users, on a sliding scale rising to
\$4\,000/year.\footnote{\url{http://arxiv.org/help/support/whitepaper}}
JISC Collections is one of these \q{tier~1} supporters, on behalf of UK
colleges and universities.

The NASA \gls{ADS} at the Smithsonian Astrophysical
Observatory preserves bibliographic information for the astronomy
literature, holds references to or copies of journal article full
texts, and curates digitised copies of older articles sometimes
unavailable from publishers.  It also curates links between these
publications and the arXiv, and between publications and data.
See~\cite{accomazzi10} for context, and some discussion of the arXiv
numbers mentioned above.

The publication paradigm represented by arXiv (and similar
smaller-scale efforts) is underpinned by the peer review processes of
journals.  However as journal subscription costs rise, journals are
progressively cancelled, in a process which may ultimately damage the
reviewing process on which the paradigm depends.  The SCOAP$^3$
consortium\footnote{\url{http://scoap3.org/about.html}} aims to break
out of this cycle by directly supporting a small number of HEP
journals, through a levy on the funding agencies which support the
field, in proportion to the share of HEP publishing they support.  In
return for this the journals will remove both subscription charges and
page charges for these journals.

\subsection{Virtual Observatories}
\label{s:vo}

\index{virtual observatory|(} 
A Virtual Observatory is an astronomical
data-sharing system, composed of a network of archives and data-access
protocols.  The goal is that the data appears to be integrated and
ideally appears to be local.

The earliest \glspl{VO} were Astrogrid in the UK, the US-VO
in the US (which became NVO and then VAO), and the Astrophysical
Virtual Observatory in Europe (which became
Euro-VO). They, along with a growing collection of
smaller national or regional VOs, formed the \gls{IVOA} in 2002.\footnote{\url{http://www.astrogrid.org},
  \url{http://www.usvao.org/}, and
  \url{http://www.euro-vo.org}; plus \url{http://www.ivoa.net}.}
The IVOA exists to broker portable network protocols for sharing data,
on the part of cooperating archives, and accessing it, on the part of
client applications.  The IVOA focuses primarily on \q{traditional} astronomy, and so has
poor coverage of solar physics and more broadly geophysics (and
certainly provides no access to GW data).  

From this has grown the more general notion of the \q{VxO},  which is \qq{[a] service that ensures that all resources from 
sub-field~$x$ are known, discoverable, and easily accessible.  It looks
to the user like a uniform data provider, but it is
virtual.}\footnote{See further commentary in
\url{http://lwsde.gsfc.nasa.gov/VxO_Report_Decadal_Survey_5_2011.pdf}}
Examples include the Virtual Solar-Terrestrial
Observatory~\cite{fox09}, HELIO, and NASA's Heliophysics Data
Environment.\footnote{\url{http://www.helio-vo.eu} and \url{http://lwsde.gsfc.nasa.gov/}}
\index{virtual observatory|)}

\subsection{Data products and proprietary periods: reifying data
  management and release}
\label{s:reification}

\index{data products|(}
A common feature of the various data styles above is the notion of the
\emph{data product}, and it seems useful to recap and stress the
salient features of this here.
\begin{statement}{Data products}
A data product is a designed and documented output of an instrument,
intended to be both archivable and immediately useful to other researchers,
by virtue of having observational artefacts removed as much as
possible.
\end{statement}
Depending on the discipline and the engineering complexity of the
instrument, data products may be anything from the raw data to a
highly processed derivative of the raw data; the ideal data product
contains all the scientifically relevant information with none of the
experimental artefacts.

Researchers are not restricted to using only data products, but it
will only rarely be necessary for them to resort to reanalysing raw
data (see the discussion on p.\pageref{s:rawdata}).

Data products correspond closely to the \q{\glspl{Information
    Package}} of the OAIS model (see \prettyref{s:oais}).  In our
experience, there tends to be little practical difference between
\glspl{SIP} and \glspl{AIP}, and where there are distinct \glspl{DIP},
they tend to be available in addition to the available SIPs and AIPs.
An exception to this is archives such as the Wide-Field Astronomy Unit at
Edinburgh,\footnote{\url{http://www.roe.ac.uk/ifa/wfau/}}\index{WFAU,
  Edinburgh} which specialises in astronomical survey science, and
develops enhanced archives (which is to say, value-added AIPs) as part
of its participation in collaborative astronomy projects.

When the various Packages differ, they tend to be regarded as
successively higher-level, as opposed to alternative, data products.

The notion of data products has a number of concrete advantages.
\begin{itemize}
\item Most immediately, the existence of a stable and documented
  output makes it easier for researchers to use and repurpose experimental
  and observational results.
\item Because the products are so central to an instrument's output,
  they, and the pipelines that produce them, are designed and costed
  at early stages of an instrument's production.
\item Researchers can produce and share software which processes
  well-defined products, possibly from more than one instrument.
\item Because they are so explicit, they form well-defined start and
  end points of discussions about interoperability between
  instruments.  Indeed, the \gls{VO} programme could be characterised
  as an extended effort to negotiate new common products which
  archives and software developers agree can be successfully generated
  (by archives) from existing AIPs.
\end{itemize}
There is of course a cost associated with the design and development
of data products, but we believe that this will in most cases be much
smaller than the costs associated with the retrospective documentation
and distribution of \adhoc\ datasets.

\index{proprietary data|(}
Another notion that is well-known in the physical sciences, but which
as far as we are aware is rare outside, is that of explicit
\emph{proprietary periods} for data.
\begin{statement}{Proprietary period}
A \q{proprietary period} is a period after data
is acquired, and therefore archived, by a shared instrument, during which it is private to the
observer or observers who requested it, and after which the data
(usually automatically) becomes public.
\end{statement}
The term \q{embargo period} would possibly be more generally
intelligible, but \q{proprietary} is conventional.
The notion is discussed
elsewhere in this document (see for example \prettyref{s:astrodata}),
but we stress it here because it usefully concretizes a number of
otherwise vague questions about data release.  

Instead of rather broad questions of the how, when, why and whether of
data management and release, we instead have questions such as \q{what
  are the data products?}, \q{whom are they documented for, and how
  expensively?}, \q{how long is the proprietary period?} or \q{what is
  the quid pro quo for this period?}\footnote{Compare the comments
  about Herschel data in \prettyref{s:herschel}.} These questions don't
magically become easy to answer, but they become a lot easier to ask,
and invite concrete answers and negotiation rather than \adhoc\ argument.

There is nothing in the notions of data products and proprietary periods which is obviously
specific to the physical sciences.  The notions have become
well-established in this area probably because it has long
experience, of necessity, of using large shared instruments which are
operated to a greater or lesser extent as services.  This is less
often the case in disciplines with more bench-scale experimental
norms, but even some areas of biology are now more often using shared
facilities, and in other disciplines, data products and proprietary
periods would become more natural, the more that preservation-aware
storage is used~\cite{factor07}.

We commend the notions of data products and proprietary periods, and the data culture they
engender, to the broader research community.  Indeed, we recommend
that  \recommendation{data managers should consider adopting the
  language of data products and explicit proprietary periods when
  designing and documenting their holdings}.
\index{proprietary data|)}
\index{data products|)}


\section{The responsibilities for data preservation}
\label{s:responsibilities}



\subsection{Visualising benefits}
\label{s:benefits}

\index{benefits|(}Why do funders wish to preserve data?
Because they perceive \emph{benefits} to that preservation.

\begin{wrapfigure}{O}{6cm}
\includegraphics{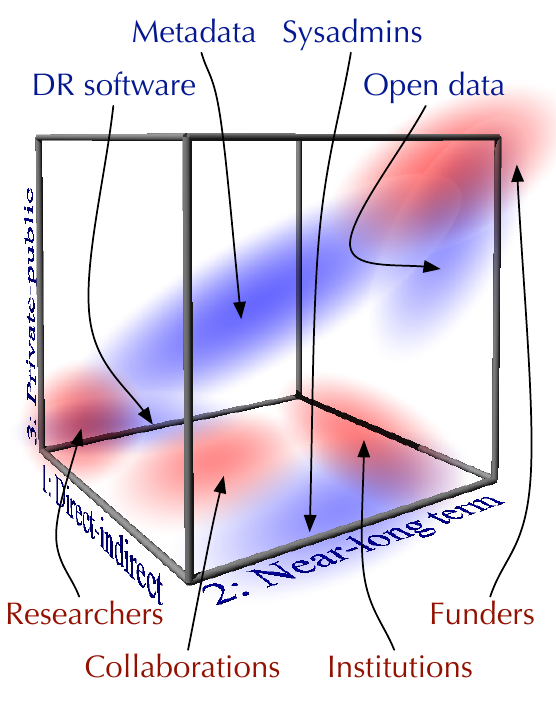}
\caption{\label{f:benefits}Visualizing benefits}
\end{wrapfigure}

Building on this truism, it seems useful to explicitly articulate
these benefits.  The JISC-funded project \gls{KRDS} (see
\url{http://www.beagrie.com/krds.php} and~\cite{beagrie10}) described
a collection of studies and tools supporting data preservation.
Amongst the KRDS innovations was a typology of \emph{benefits},
describing three dimensions: direct to indirect, near- to long-term
and public to private.  In a slight extension to the work in KRDS, we can
take the notion of \q{dimensions} perfectly literally, assign any
particular benefit to a position along each of the three axes, and
plot the result in a three-dimensional space; see
\prettyref{f:benefits}.

In this figure we identify four
benefits which might be associated with a big-science project\dash namely
the existence of data-reduction software, good metadata, the provision
of open data and the existence of system administrators\dash and we sketch
the approximate volumes they might occupy along the three axes (in
blue).  On the same diagram, we can indicate (in red) the approximate areas of
interest of four sample stakeholders. 

In the example here, \q{sysadmin support} can be seen as an indirect
benefit to researchers, typically private to an institution, but
creating value in the near- and long term; it is therefore spread
along the \q{near-long term} axis, but at one extreme of the other two
dimensions.  We can put on the same
diagram the approximate areas of interest of various research
stakeholders.  For simplicity, we are here conceiving of individual
researchers as selfish and short-termist, though the same researchers
will have long-term interest when they have a collaboration or institutional hat on,
and indirect public interests in the long-term health of their
discipline when they are serving on a funding council grants panel;
below we will take the term \q{funders} to refer both to the officials
of funding bodies, acting as proxies for the wider interests of
society, and to members of the research community discharging service roles.

We should not take this diagram too literally\dash it is not clear that
the axes are independent, and the extent and even the gross positions
of the various interests and benefits are debatable.  The
diagram is nonetheless thought-provoking.  For example, it visually predicts that much of
the research community is not particularly interested in \q{open
  data}\footnote{(unless it's \emph{other people's} open data, of
  course)} and only incompletely interested in \q{good metadata}
(in-collaboration researchers care when a dataset was acquired,
because they need that information to perform their analyses, but they
have little interest in dissemination and licensing metadata, for
example, because that is the long-term concern of funders and their proxies).  We can
therefore naturally conceive of the funders taking the role of the
conscience of a discipline, worrying about long-term imponderables so
that individual researchers don't have to.  It follows from this, that
the open data case made to funders, for example, will be an
institutionally self-interested one, but that the case made to 
researchers must be qualitatively different, and be either pragmatic
(\q{you must care because your funders care}) or high-minded (\q{your
  socio-cultural duty is\dots}).  Neither of these is a poor argument,
nor indeed a cynical one, but we are acknowledging here that, to a busy and
distracted researcher, the self-interest argument in isolation may have little purchase.
\index{benefits|)}

\subsection{The case for open data}
\label{s:openness}
\index{open data|(} Internationally, there is a push towards such
\gls{data sharing} in the more general context of scholarly research
(see for example~\cite{Arzberger2004} or~\cite{ruusalepp08}).  The
most explicit statement here is in the \gls{NSF}'s GC-1
document~\cite{nsf-gc1}, which in section~41 states that \qq{[NSF]
  expects investigators to share with other researchers,  at no more
  than incremental cost and within a reasonable time, the data,
  samples, physical collections and other supporting materials created
  or gathered in the course of the work. It also encourages grantees
  to share software and inventions or otherwise act to make the
  innovations they embody widely useful and usable.} This is
reiterated in almost the same words in their 2010 data sharing policy~\cite{nsf11}.
They additionally require a brief statement, attached to proposals, of
how the proposal would conform to NSF's data-sharing policy.

\gls{STFC}, in common with the other UK research councils, requires
that \qq{the full text of any articles resulting from the grant that
  are published in journals or conference proceedings [\dots] must be
  deposited, at the earliest opportunity, in an appropriate e-print
  repository}\footnote{\url{http://www.scitech.ac.uk/rgh/rghDisplay2.aspx?m=s&s=64}};
it has not yet made any corresponding statement on data releases.

The year 2009 saw some excitement (relating to the incident inevitably
labelled \q{climategate}, and to some other data-release
disputes\footnote{\longurl{http://www.guardian.co.uk/~environment/~2010/~apr/~20/~climate-~sceptic-~wins-~data-~victory}})
related to the management and release of climate data.\index{climate
  data} This illustrated the political and social significance of some
science data sets; the contrast between what scientists know, and the
public believes, to be normal scientific practice; and some of the
issues involved in the generation, ownership, use and publication of
data.\footnote{UEA's Climate Research Unit is a partner in the ACRID
  project, also funded by the JISC \gls{MRD}
  programme: \url{http://www.cru.uea.ac.uk/cru/projects/acrid/}}
The cases during that year
illustrate a number of complications involved in data releases.
\begin{enumerate}
\item Data is often passed from researchers or groups directly to
others, across borders, with no general permission to distribute it
further.
\item Data collection may be onerous, and the result of significant
professional and personal investments.
\item Raw data\index{raw data!utility} is generally useless without the more
or less significant processing which cleans it of artefacts and makes
it useful for further analysis.
\item However not all disciplines have the clear notion of published data
products\index{data products} which is found in astronomy and which is
implicit in the OAIS notion of archival deposit.\label{item:dataproducts}
\item Science is a complicated social process.\footnote{The last point is
  simultaneously obvious and deeply intricate.  Unpacking it would
  distract us here, but there is further discussion, in a very apposite
  historical context, in~\cite{collins04}, elaborated in~\cite{collins07}.}
\end{enumerate}
In science, we preserve data so that we can make it available
later. This is on the grounds that scientific data should generally be
universally available, partly because it is usually publicly paid for,
but also because the public display of corroborating evidence has been
part of science ever since the modern notion of science began to
emerge in the 17th century (\textsc{ce})\dash witness the Royal Society's
motto, \q{nullius in verba}, which the Society glosses as \q{take
  nobody's word for it}. Of course, the practice is not quite as
simple as the principle, and a host of issues, ranging across the
technical, political, social and personal, complicate the social,
evidential and moral arguments for general data release.

The arguments \emph{against} general data releases are
practical ones: data releases are not free, and may have significant
financial and effort costs (cf \prettyref{s:preservation-costs}).\index{costs}
Many of these costs come from
(preparation for) data preservation, since it is formally archived data
products that are the most naturally releasable objects: releasing raw or low-level
data \emph{may} be cheap, but may also have little value, since raw
underdocumented datasets are likely to be
useless; or more pessimistically they may have a negative value, if they
end up fostering misunderstandings which are time-consuming to counter
(this point obviously has particular relevance to politicised areas such as
climate science).
In consequence of this, the \q{open data question} overlaps with the
question of data preservation\dash if the various costs and sensitivities
of data preservation are satisfactorily handled, then a significant
subset of the practical problems with open data release will promptly
disappear.  We discuss the data preservation question below, in
\prettyref{s:case-data-preservation}.

It seems worth noting, in passing, that the physical sciences broadly
perform better here than other disciplines, both in the technical
maturity of the existing archives and in the community's willingness
to allocate the time and money to see this done effectively.

What all this indicates is that there is a need for an
explicit framework for discussing the pragmatics of open data (cf
point~\ref{item:dataproducts} above).  We can
go further and suggest (it is almost a Recommendation) that the OAIS\index{OAIS}
model's notion of an \gls{AIP}, and its
reflection in the notion of a \emph{data product}\index{data products}
should be central to this discussion.
\index{open data|)}

\subsection{The case for data preservation}
\label{s:case-data-preservation}

The case for data preservation in astronomy was implicitly made in
\prettyref{s:astrodata}: as an observational science, much astronomy
data is repeatable, but there are important cases where what is being
observed is a slow secular change, or some unpredictable (usually
ultimately explosive) event; sometimes data can be opportunistically
reanalysed to extract information distinct from the information the
observation was designed for.  Astronomical data is potentially useful
\emph{and} usable almost indefinitely.  Thus there is a reasonable
expectation that the data can be and will be exploited by unknown
astronomers, far into the future.

HEP data is somewhat different (as noted in \prettyref{s:hepdata}).
As an experimental science, it is generally very much in control of
what it observes, and is able to design experiments of considerable
ingenuity, in order to make measurements of exquisite discriminatory power.
A consequence of this is firstly that HEP experiments have a much
stronger tendency to become obsolete with each technological
generation, and secondly that the complication of the apparatus makes
it hard to communicate into the future a level of understanding
sufficient to make plausible use of the data.  Experimental apparatus
will generally be understood better and better as time goes on (this
is also true of satellite-borne detectors in astronomy), so that data
gathered early in an experiment will be periodically reanalysed with
increased accuracy.  However this understanding is generally not
preserved formally, but is pragmatically communicated through wikis,
workshops, word of mouth, configuration and calibration files, and
internal and external reports.  Even if all of the tangible records
were magically preserved with complete fidelity, and supposing that
the more formal records do contain all the information required to
analyse the raw data, an archive would still be missing the
word-of-mouth information which a new postgrad student (for example)
has to acquire before they can understand the more complete
documentation.  We can think of this as a \q{bootstrap problem}.  In
OAIS terms, the \gls{Representation Network}
for HEP data is particularly intricate, and while the
\gls{Representation Information} nearest to the \gls{Data Object} may
be complete, it may be infeasible to gather the Representation
Information necessary to let a naive researcher make sense of it.  The
\gls{Designated Community} for HEP data may therefore be null in the
long term.

This sounds pessimistic, but~\cite{south11} describes a number of
scenarios in which HEP data can and should be reanalysed some decades
after an experiment has finished, and describes ongoing work on the
development of consensus models for preserving data for long enough to
enable such post-experiment exploitation.  This provides a strong case
for a style of preservation somewhat different from the astronomical
one.  What these models have in common is a commitment of staff to
actively conserve and continuously exploit the data.  This
post-experiment staff can therefore 
be conceived as a form of walking Representation Information so that,
while they are still involved, the data might have a Designated
Community which corresponds to those individuals in a position to
undertake an extended apprenticeship in the data analysis (this model
is further discussed on p.\pageref{s:hepdmpplans}).

GW data is, as usual, somewhere between these two extremes.
As astronomy, the GW data consists of unrepeatable measurements which
will potentially be of value to astronomers well into the future; as a
HEP-style experiment it makes those measurements using two or three
generations of highly sophisticated apparatus, each generation of
which will improve on the sensitivity of its predecessors by orders of
magnitude.  An additional feature, however, is that no-one has ever
convincingly detected a gravitational wave, though there have been
repeated claims of detection in the past, so that the first claims by
LIGO or \gls{aLIGO} will be scrutinized particularly closely.

Finally, and as noted in \prettyref{s:openness}, if data is well
archived, then most of the pragmatic objections to opening that data
do not apply.  Thus, to the extent that general data release is a good
in itself, it is a further argument in favour of a well supported archive.

\subsection{Should raw data be preserved?}
\label{s:preserve-raw}

\index{raw data!preservation|(}
In the data-preservation world, there is often an automatic expectation that \q{everything
  should be preserved}, so that an experiment can be redone, results reanalysed, or an
analysis repeated, later.  Is this actually true?  Or if it is at
least desirable, how much effort should be expended to make it true?
This question is implicit in, for example, the discussion of software
preservation\index{software preservation} in
\prettyref{s:sw-preservation}.

When a physical experiment is set up and working, it is usual to avoid
tinkering with it as much as possible, to avoid any unexpectedly
significant change.  That is, even with a small-scale lab-bench
experiment, it is accepted that not everything can be effectively
documented, and that an experiment might not be immediately replicable
purely from published information (cf \cite[ch.35]{collins04} and \prettyref{s:socscidata}).  This
expectation (or rather, lack of expectation) is also true of
larger-scale experiments, which might be financially, professionally
or, at the largest scales, politically infeasible to
replicate.\footnote{It is because very large-scale experiments are
  impossible to replicate, and even hard for an external reviewer of
  an article to criticise meaningfully, that large collaborations
  submit their publications to extremely scrupulous internal review.
  See \url{http://stuver.blogspot.com/2011/03/big-dog-in-envelope.html}
  for a post-mortem account of such a review.}
Perhaps this attitude should extend to other aspects of the
experimental process.

In many cases, the pipeline for reducing raw data seems to fall into
this category: it encodes hard-to-document information, but is itself
hard to document, hard to use, and unlikely ever to be reused in fact.
If this software is not preserved, then the raw data is effectively
unreadable, which means there is no case for preserving it.
There is therefore a case that at least some details of the
experimental environment\dash digital as well as physical\dash are not
reasonably preservable, and that as a result
little effort should be expended on preserving them.

It is data products\index{data products} that make raw data less
necessary.  It is feasible to document the scientific meaning of data
products, and the community expects that a project will provide this
documentation as part of the publication of the products (indeed, it
the documentation that makes these
\emph{products} rather than just a casual data snapshot).  The data
products allow researchers to dig beneath the conclusions of a
particular article (or indeed the contents of a higher-level data
product), and to criticise and build on what they find there.
Higher-level products are the result of higher-level scientific
judgements, and it is normal for these to be regenerated by
researchers other than the originators, either using their own
software or the originators' pipelines.  These later-stage pipelines
are more formally supported by projects, which involves making them
reasonably portable, so that they are both easier to preserve as well
as being more valuable objects of preservation.

We should stress that we are not advocating deliberately deleting raw
data, and its associated pipelines\dash it \emph{might} be useful, and it
\emph{might} be usable\dash but simply noting that one should not
overstate its value.
\index{raw data!preservation|)}

\subsection{OAIS: suitability and motivation}
\label{s:oais-motivation}

In \prettyref{s:oais}, we provide an overview of the OAIS model, and
describe how it relates to astronomical data.

The OAIS standard is formally a product of the \gls{CCSDS},\footnote{\url{http://www.ccsds.org}} and with
this in its lineage it is quite naturally matched to the data
management problems of the physical sciences.  Essentially all the
explicit and implicit assumptions of the OAIS standard are true in the
area we are studying: the data producer (a satellite or a detector) is
usually obvious, the various \gls{Information Package}s (or data
products) well understood, and the \gls{Designated Community}
easily identified.\footnote{There is no contradiction here with the
  remarks in \prettyref{s:socscidata} about the difficulty of
  describing the \gls{Designated Community} of science archive users.
  It is easy to name a science Designated Community, but it may be hard
  to describe ahead of time what those community members can be
  expected to know.  A social science archive may have an unpredictably broad range
  of ultimate users, but using the archive will need
  little specialist knowledge; in contrast a particle physics dataset will
  probably be of interest only to particle physicists, but the normal
  education of such a physicist three decades hence, and thus the
  content and extent of the specialised Representation Information that
  Community will need, might be very hard to guess at.}

The motivation for a digital preservation standard, as discussed in
the OAIS standard itself~\cite[\S2]{std:oais}, is that digital
preservation represents a double problem: (i)~digital information is
intrinsically harder to preserve than traditional information, which
is capable of sitting on a shelf in a well-understood and intelligible
format, and mouldering at a well-understood and graceful rate; and
(ii)~more and more organisations are producing digital information
\emph{and} are implicitly expected to archive their own material.
This means that these non-specialist archives have a complicated task
to perform, which is potentially at odds with the daily urgencies of
their main business.

This \emph{appears} to mean in turn (and in JISC contexts it is often
taken to mean in practice) that these
organisations need as much detailed and prescriptive help as possible,
ideally devolving their archive responsibilities to a central
discipline- or funder-specific archive, to the extent possible while
respecting the low-level complications and friction alluded to in
\prettyref{s:socscidata}.  This is not the model which is appropriate for big-science datasets.

\subsection{What should big-science funders require, or provide?}
\label{s:funder-requirements}

We have described several common features of big-science data
management in \prettyref{s:styles}. and we have outlined some
particular contrasts with other communities in
\prettyref{s:socscidata}.  As noted in
\prettyref{s:curatingbigscience}, our focus here is on STFC's
strategically funded projects, rather than the smaller projects funded
by individual research grants.

Big-science data sets are generally intimately coupled to solutions to
leading-edge technical challenges, and cannot usefully be regarded as
incremental changes to previous solutions.  This, coupled with the
general availability of extensive technical expertise within such
communities, means that any generic solution is very unlikely to be
appropriate, and that it is both reasonable and feasible to require custom
archiving solutions for such projects.  There is no
\emph{recipe} for data preservation on this scale, and all that can be
hoped for is a structured approach to a custom solution.  Having said
this, not even the most innovative science experiments are so
completely \emph{sui generis} that they warrant a data preservation
approach which is reimagined from scratch.  It is therefore wasteful
to ignore the considerable intellectual investments in the OAIS model,
the growing penumbra of commentaries on and developments of it, and the
minor industry of validation and auditing efforts related to it.

We are therefore led to the conclusion that the most effective overall
strategy for effective data management in the large-scale experimental
physical sciences is that \recommendation{funders should simply
  require that a project develop a high-level DMP plan as a suitable
  profile of the OAIS specification~\cite{std:oais}}.\index{OAIS}
This profile should be detailed enough to require negotiation with the
funder and with the experiment's community, but can leave many
of the implementation details to the good engineering judgement of the
project's management.  We believe the LIGO \gls{DMP} plan\index{LIGO!DMP}~\cite{anderson11} can
be taken to be exemplary in this regard.

Big-science projects have the technical skills, the management
structures, and the budgets to take on such a task, and to deliver a
custom archive which can be shown to meet identified goals.
We recommend that \recommendation{funders should support projects in creating
  per-project OAIS profiles which are appropriate to the project and
  meet funders' strategic priorities and responsibilities}.  

The discussion in \prettyref{s:aida} suggests that one result of the
development of an OAIS-based \gls{DMP} plan is that the resulting plan is
explicit enough to generate useful deliverables, and to benefit from
the growing interest in OAIS \q{validation}.

We suggest the following specific funder actions.
\begin{itemize}
\item Actively engage with projects to help them develop an OAIS
  profile.  This will include overview literature, including the OAIS
  specification, tutorial reports such as~\cite{lavoie04}, and commentary
  such as~\cite{rosenthal05}, or perhaps specialised workshops if
  necessary.  These are high-level introductions,
  rather than pro\-ce\-dure-based tick-lists.
\item Develop or support expertise in criticising and validating such
  OAIS profiles.  For example, the CASPAR consortium (see for
  example~\cite{caspar-evaluation09}) has developed strategies for
  detailed validation of projects' claims about long-term data
  migration.  Similar work\dash for example validating a project's
  assumptions about its \gls{Designated Community}\dash would reassure
  the wider community that the archive design is likely to achieve its
  goals for the future.
\end{itemize}

The first of these is reasonably straightforward, consisting of little
more than gathering resources.
The second is a longer-term project which may require some expertise
to be built up and supported at a funder-supported facility (such as
RAL, in the UK), or through liaison with the \gls{DCC}.

A corollary of this more active engagement is that funders must
financially support the preservation work they require.  See \prettyref{s:preservation-costs}.


\section{The practicalities of data preservation}
\label{s:practicalities}






\subsection{Modelling the archive}

\subsubsection{The OAIS model}
\label{s:oais} 

\index{OAIS|(}
We introduce here the main concepts of the OAIS model.  Full details
are in~\cite{std:oais} with a useful introductory guide
in~\cite{lavoie04} and some discussion in the LSC context in \cite{anderson11};
the OAIS motivation is further discussed in
\prettyref{s:oais-motivation}.

\begin{figure}
\includegraphics[width=\textwidth]{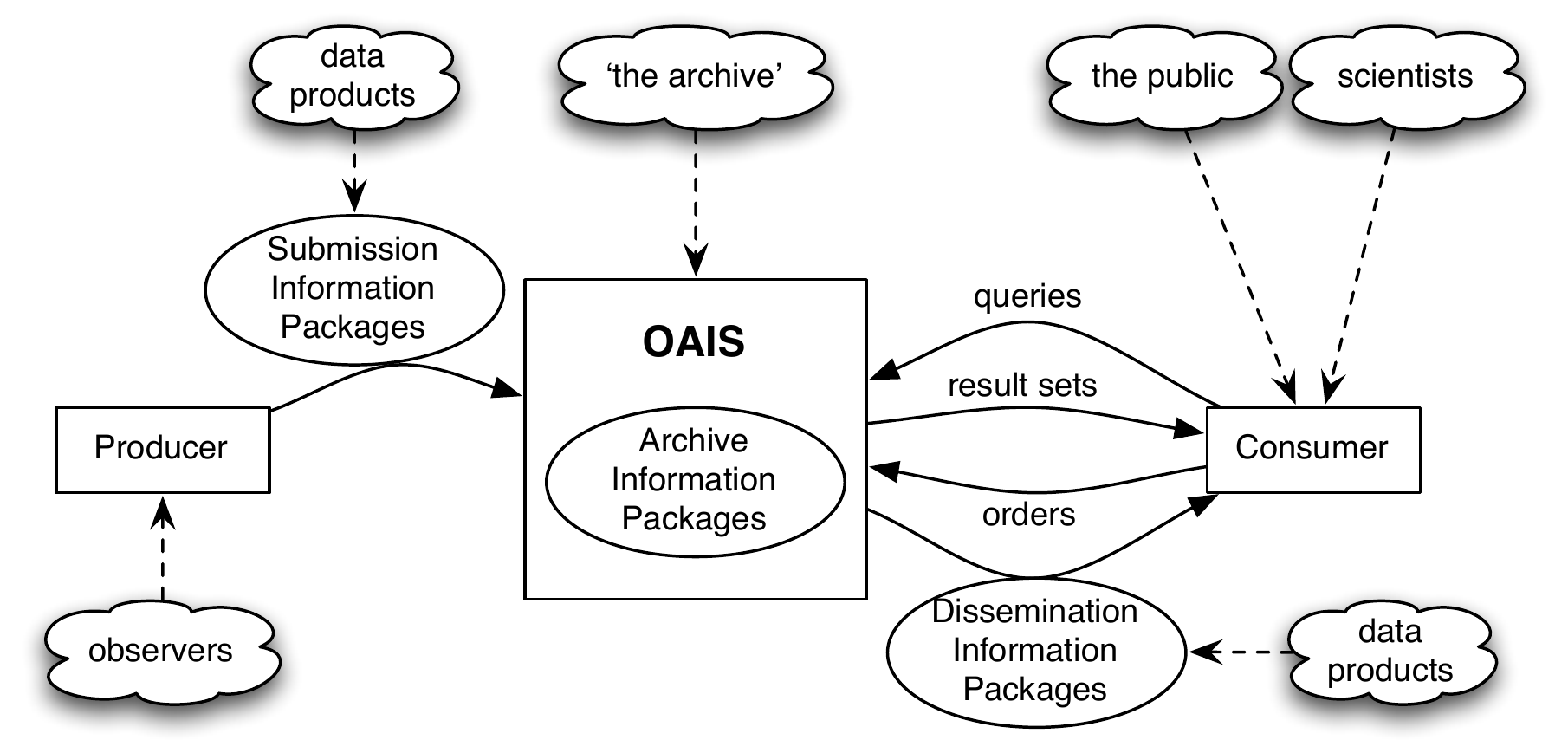}
\caption{\label{f:oais-annotated}The highest-level structure of an
  OAIS archive, annotated with the corresponding labels from
  conventional astronomical practice (redrawn from~\cite[Fig.~2-4]{std:oais}).
  The dissemination data products will typically be the same as the
  submitted ones, but archives can sometimes create value-added ones
  of their own.}
\end{figure}

The term \emph{OAIS} stands for an \emph{Open Archival Information
  System}.  The word \q{open} is not intended to imply that the archived
data is freely available (though it may be), but instead that the
process of defining and developing the system is an open one.
The principal concern of an OAIS is to
preserve the usability of digital artefacts for a pragmatically
defined long term.  An OAIS is not only
concerned with storing the lowest-level \emph{bits} of a digital
object (though this part of its concern, and is not a trivial problem), but with storing enough
\emph{information} about the object, and defining an adequately
specified and documented \emph{process} for migrating those bits from
system to system over time, that the information or knowledge
those bits represent can be retrieved from them at some indeterminate
future time.  The OAIS model can therefore be seen as addressing an
administrative and managerial problem, rather than an exclusively
technical one.

The OAIS specification's principal output is the \emph{OAIS reference
  model}, which is an explicit (but still rather abstract) set of
concepts and interdependencies which is believed to exhibit the
properties that the standard asserts are important
(\prettyref{f:oais-annotated}).
The OAIS model
can be criticised for being so high-level that \qq{almost any system
  capable of storing and retrieving data can make a plausible case
  that it satisfies the OAIS conformance
  requirements}~\cite{rosenthal05}, and there exist both efforts to
define more detailed requirements~\cite{rosenthal05}, and efforts to
devise more stringent and more auditable assessments of an OAIS's
actual ability to be appropriately responsive to technology
change~\cite{caspar-evaluation09}.

An OAIS archive is conceived as an entity which preserves objects
(digital or physical) in the \gls{Long Term}, where the \q{Long Term} is
defined as being long enough to be subject to technological change.  The
archive accepts objects along with enough \gls{Representation
  Information} to describe how the digital information in the object
should be interpreted so as to extract the information within it (for
example, the FITS specification is Representation Information for a
FITS file).   That Information may need further context\dash for example,
to say that a file is an ASCII file requires one to define what ASCII
means\dash and the collection of such explanations turns into a
\gls{Representation Network}.  This information is all submitted to the archive in the form
of a \gls{SIP} agreed in some more or less
formal contract between the archive and its data producers.

Once the information is in the archive, the long-term responsibility
for its preservation is \emph{transferred} from the provider to the archive,
which must therefore have an explicit plan for how it intends to
discharge this.

The archive distributes its wares to Consumers in one or more
\glslink{Designated Community}{Designated Communities}, by transforming them, if necessary,
into the \gls{DIP} which corresponds to
a \q{data product}.  The members of the Designated Community are those
users, in the future, whom the archive is designed to support.  This
design requires including, in the \gls{AIP},
Representation Information at a level which allows the Designated
Community to interpret the data products \emph{without ever having met
  one of the data Producers}, who are assumed to have died, retired,
or forgotten their email addresses.

The OAIS model originated within the space science community, so it
can be mapped to the physical science data of the GW community without
much violence.
\index{OAIS|)}

\subsubsection{The DCC Curation Lifecycle model}
\label{s:dcc-lifecycle}
\index{DCC lifecycle|(}
The OAIS model is on the face of it a linear one, and suggests that data is created,
then ingested, then preserved, and then accessed, in a process which has a clear
beginning and end.  This is compatible with the observation that
one point of archiving data is to reuse or repurpose it, creating new archivable
data products in turn, but this longer-term cycle remains only implicit in
the model.  The OAIS model is therefore very usefully explicit about
those aspects of archival work concerned with long-term preservation,
but its conceptual repertoire is such that a discussion framed by it
runs the risk of underemphasizing the range of roles a data
repository has, or even of marginalising it.

\begin{figure}
\begin{centering}
\includegraphics[width=0.75\columnwidth]{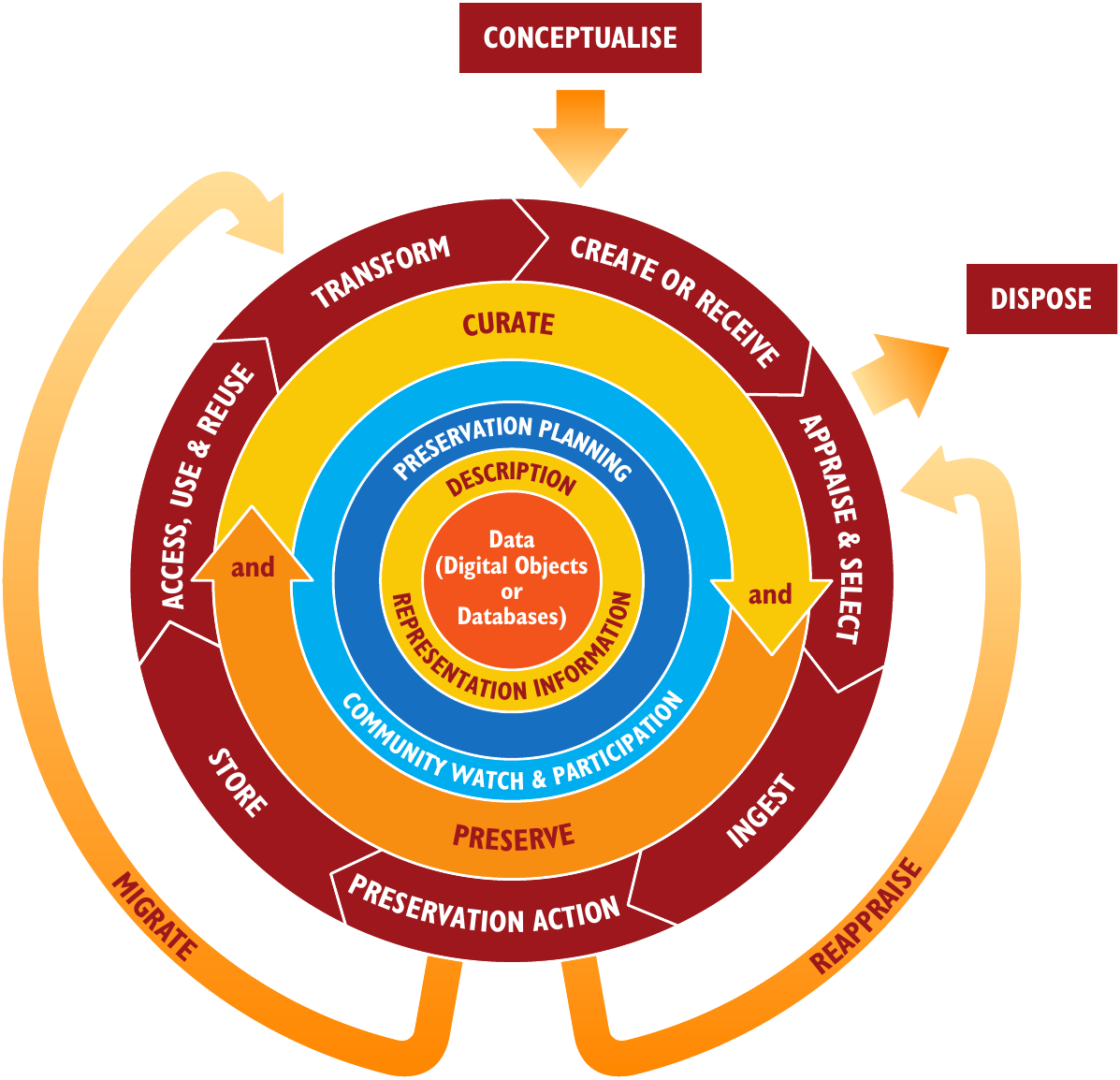}\\
\end{centering}
\caption{\label{f:dcc-lifecycle}The DCC lifecycle model, from \cite{dcc10}}
\end{figure}

In contrast,\footnote{We thank Dorothea Salo, of the University of
  Wisconsin library, for emphasizing to us the useful applicability of
  the DCC model to the case of big science data, and Angus Whyte, for
  elaborating the contrasts between the DCC and OAIS models.} the
\gls{DCC} has produced a lifecycle model~\cite{dcc10} (\prettyref{f:dcc-lifecycle})
which stresses that data creation, management, and reuse are part of a
cycle in which preservation planning, for example, can naturally
happen before data creation as well as after it; and in which data can
be appraised, reappraised, and possibly disposed of if it becomes
obsolete.  It therefore makes explicit both the short- and long-term
cycles in the flow of active research data, and it emphasizes the
active involvement of data curators in maintaining that cycle.

Cycles of use and re-use are not the only links between datasets.  As
discussed in~\cite{lee07}, one digital object can also provide context
for another, in a variety of ways.  To some extent this remark
rediscovers the notion of the OAIS \gls{Representation Network}, and this
in turn prompts us to stress that although we have contrasted OAIS and
DCC here, they are not in competition: OAIS is concerned with the
creation and management of a working archive with gatekeepers
and firm goals; the DCC model is concerned with the location of the
archive in the wider intellectual context.

The DCC model is immediately compatible with the observation, in
\prettyref{s:preservation-costs} below, that HEP and GW archives
effectively avoid some preservation costs by seeing long-term
preservation as only part of the role of a data repository.  Accepting
data, making it available as working storage, transforming it into
immediately useful forms, or appraising (possibly regenerable)
datasets whose storage costs outweigh their usefulness, all give the
archive a familiarity with the data, and the researchers a familiarity
with the archive, which means that the decision to select certain data
for long-term preservation is potentially more easily reached, more
easily defended and more easily funded, than if the archive is
conceived as a cost-centre bucket bolted on the side of the project.
This appears to be borne out by the LIGO experience, in which the
new \gls{DMP} plan was developed and successfully argued for by the same personnel who were
long responsible for the design and management of the data management
system on which everyone's daily work depends.
\index{DCC lifecycle|)}

\subsection{Software preservation}
\label{s:sw-preservation}
\index{software preservation|(}As discussed in, for example, \prettyref{s:gwdata}, there is often a
substantial amount of important information encoded in ways which are
only effectively documented in software, or software configuration
information.  There is therefore an obvious case for preserving this
software (though note the caveats of \prettyref{s:preserve-raw}).

Preservation of a software pipeline requires preserving the \gls{pipeline} software
itself, a possibly large collection of libraries the software depends
on, the operating system (OS) it all runs on, and the configuration
and start-up instructions for setting the whole thing in motion.  The
OS may require particular hardware (CPUs or GPUs), the software
may be qualified for a very small range of OSs and library versions,
and it may be hard to gather all of the configuration information
required (there is some discussion of how one approaches this problem
in for example~\cite{south11}).  It is not certain that it is
necessary, however: if the data products are well-enough described,
then re-running the analysis pipeline may be unnecessary, or at least
have a sufficiently small payoff to be not worth the considerable
investment required for the software preservation.  We feel that, of
the two options\dash preserve the software, or document the data products\dash
the latter will generally be both cheaper and more reliable as a way
of carrying the experiment's information content into the future, and
that this tradeoff is more in favour of data preservation as we
consider longer-term preservation.

This last point, about the changing tradeoff, emphasizes that the two
options are not exclusive: one can preserve data \emph{and} preserve
software, and the JISC-funded Software Sustainability
Institute\footnote{\url{http://www.software.ac.uk/}} provides a
growing set of resources which provide guidance here.  However the
solutions presented generally focus on active curation, in the sense
of preserving software through continuing use and maintenance.  This
can be successful\footnote{The UK Starlink project provided
  astronomical software.  It ran from 1980 to 2005, when it was
  rescued from oblivion by being taken up by the UK Joint Astronomy
  Centre \ifxetex Hawai‘i\else Hawai`i\fi.  The current distribution includes still-working
  code from the 80s.  The Netlib and BLAS libraries have components
  which date from the 70s.}, and is the approach implicit
in~\cite{south11}, but it seems brittle in the face of significant
funding gaps, and would not deal well with the case where a software
release is deliberately unused, for example because it has been
superseded.\index{software preservation|)}

\subsection{Data management planning}

\subsubsection{DMP in space}

As one might expect, both NASA and ESA have formalised \gls{DMP} plans.

\gls{NASA}'s \gls{NSSDC}\footnote{\url{
    http://nssdc.gsfc.nasa.gov}} has led NASA's data planning since
the mid-80s.  It was initially the NSSDC which negotiated a Project
DMP plan with missions, but since the 1990s this has become the
responsibility of the NASA
\gls{PDS}\footnote{\url{http://pds.nasa.gov/}}.  The NSSDC's data
retention policy\footnote{\url{http://nssdc.gsfc.nasa.gov/nssdc/data_retention.html}}
describes what categories of data product should be retained
indefinitely, and the PDS provides resources to mission planners on
the processes and
tools\footnote{\url{http://pds.nasa.gov/tools/index.shtml}} for
preparing data for preservation.\footnote{We are grateful to Paul
  Butterworth of NASA for helpful advice here.}

\gls{ESA}'s Planetary Science
Archive\footnote{\url{http://www.rssd.esa.int/index.php?project=PSA&page=about}}
\qq{provides expert consultancy to all of the data producers
  throughout the archiving process. As soon as an instrument is
  selected, PSA begin working with the instrument team to define a set
  of data products and data set structures that will be suitable for
  ingestion into the long-term archive.}  The ESA archive is by design
compatible with the PDS.

\subsubsection{Current and future DMP in the LSC}
\label{s:stuart-and-roy}
\index{LIGO!DMP|(}

The current \gls{LIGO} \gls{DMP} plan~\cite{anderson11},
discusses DM planning with an emphasis on the preparations for the
eventual public data release.

The LIGO DMP plan proposes a two-phase data release scheme, to come
into play when \gls{aLIGO} is commissioned; this was prepared at the
request of the \gls{NSF}, developed during 2010\range11, and will be
reviewed yearly.

The plan documents the way in which the consortium will make LIGO data
open to the broader research community, rather than (as at present)
only those who are members of the LSC.  This document describes the
plans for the data release and its proprietary periods, and outlines
the design, function, scope and estimated costs\index{costs} of the
eventual LIGO archive, as an instance of an OAIS model.  This is a
high-level plan, with much of the detailed implementation planning
delegated to partner institutions in the medium term.

In the first phase, data is released much as it is at present:
validated data will be released when it is associated with detections,
or when it is related to papers announcing \emph{non}-detections (for
example, associated with another astronomical event which might be
expected or hoped to produce detectable GWs).  In the second phase\dash
after detections have become routine, and the LIGO equipment is acting
as an observatory rather than a physics experiment\dash the data will be
routinely released in full: \qq{the entire body of gravitational wave
  data, corrected for instrumental idiosyncrasies and environmental
  perturbations, will be released to the broader research
  community. In addition, LIGO will begin to release near-real-time
  alerts to interested observatories as soon as LIGO \emph{may} have
  detected a signal.}  This second phase will begin after LIGO has
probed a given volume of space-time (see \cite[ref 7]{anderson11}),
\emph{or} after 3.5 years have elapsed since the formal LIGO
commissioning, whichever is earlier.  Alternatively, LIGO may elect
to start phase two sooner, if the detection rate is higher than
expected.

In phase two, the data will have a 24-month proprietary period.\index{proprietary data}

The DMP plan describes three (OAIS) \glslink{Designated Community}{Designated Communities}.
Quoting from~\cite[\S1.5]{anderson11}, the communities are as follows.
\begin{itemize}
\item LSC scientists: who are assumed to understand, or be responsible for, all the complex details of the LIGO data stream.
\item External scientists: who are expected to understand general
  concepts, such as space-time coordinates, Fourier transforms and
  time-frequency plots, and have knowledge of programming and
  scientific data analysis. Many of these will be astronomers, but
  also include, for example, those interested in LIGO's environmental
  monitoring data. 
\item General public: the archive targeted to the general public,
will require minimal science knowledge and little more computational
expertise than how to use a web browser. We will also recommend or
build tools to read LIGO data files into other applications.
\end{itemize}

The LIGO \gls{DMP}  plan is, we believe, a good example of a plan for a project of
LIGO's size: it is specific where necessary, it was negotiated with the
project's funder (NSF) so that it achieved their goals, and it went
through enough iterations with the broader LIGO community (the agreed
version in \cite{anderson11} is version~14) that its authors could be
confident it had their approval, and that the community was
comfortable with what the DMP plan was proposing.  The document has a
strong focus on the LIGO data release criteria, since this was
the most immediate concern of both the funder and the project, but it
systematically lays out a high-level framework for future data preservation,
guided by the OAIS functional model.
\index{LIGO!DMP|)}

\subsection{Data preservation costs}
\label{s:preservation-costs}

There is a good deal of detailed information, and some modelling, of
the costs of digital preservation.\index{costs|(} The KRDS2 study
\cite[\S\S6\&7]{beagrie10} includes detailed costings from a number of
running digital preservation projects, in some cases down to the level
of costings spreadsheets.  The LIFE${}^3$ project has also developed
predictive costings tools~\cite{hole10}, and the PLANETS project
(\url{http://www.planets-project.eu/}) has generated a broad range of
materials on preservation planning, including costing studies.

Although there is a broad range of preservation projects surveyed in
the KRDS report, there are numerous common features.
Staff costs dominate hardware costs, and scale only very weakly with
archive size.  The study also notes that acquisition and ingest costs
are a substantial fraction (70\range80\%) of overall staff costs, but also
scale very weakly with archive size.  These are relatively small
archives, generally below a few~\si{TB} in size, where ingest is a
significant component of the workload.  In this report we are
interested in archives three or four orders of magnitude larger than
this where (as discussed below) ingest may be cheaper, but in broad
terms, it appears still to be true that staff costs dominate hardware
costs at larger scales, and scale only weakly with archive
size.

Parenthetically, notice that the above discussion prompts the question
\q{what is the size of an archive?}  The number of bytes it consumes
is an obvious and readily available measure, but may not be
particularly meaningful in this context.  The number of items (such as
interview transcripts, images or database rows) may be a better
measure, and still objectively identifiable, archive by archive.  If
there were some measure of abstract information content, we speculate
that this is what would scale most straightforwardly with the effort
required for quality control and metadata curation, and hence with
staff effort.  We hesitate to ask what such a measure might be, in
case the answer is \q{citation analysis}.

The lack of scaling with size, even when an archive progressively
grows in size, seems to suggest that it is an archive's \emph{initial} size
(in the sense of small, medium or large, for the time) that largely
governs the costs.

We were given access to confidential figures for the development and
operations of a mid-to-large size astronomy archive (of order \SI{10}{TB} of
relational data and \SI{100}{TB} of flat file data), developed by an
experienced archive site.  The archive software and system development
cost 25\range30 staff-years of effort: the bulk of this was for the
core database system, but between a quarter and a
third was for software to support ingest and the generation of data
products.  The organisation budgets around 3\,FTEs for operation of
this archive, which includes ingest, quality control and helpdesk
support (this is an estimated fraction of an operations team covering
several archives at the same site, so there may be some economies of
scale).  About a quarter of the annual operating budget is spent on
hardware.

The \gls{ESO} data archive manages data
from multiple ESO facilities;\footnote{We are most grateful to Fernando
  Comer\oacute n, of ESO, for sharing these figures.} it shares space with the
still-developing ALMA archive, but the figures below do not include
ALMA.  The archive is based on spinning disks backed by a tape library
(for further details, see~\cite{eglitis09}).  It currently holds
\SI{190}{\tera\byte}, increasing at around
\SI{7}{\tera\byte\per\mth}.  The hardware costs average around
\SI{330}{\kilo\eur\per\yr}, which includes hardware replacement and
data migration, and which has remained flat for some years, despite the
slowly increasing data volumes.  Running costs amount to
\SI{55}{\kilo\eur\per\yr} (some smaller systems account for part of
this), and licences, networks and other consumables account for about
\SI{30}{\kilo\eur\per\yr}.  Manpower costs come to 4~FTEs of ESO staff
plus around \SI{270}{\kilo\eur\per\yr} of outsourced staff.  Neither
hardware nor software costs appear to scale with data volume, with some
cost elements even dropping as the archive moves to completely on-line
data distribution.

There is some discussion of the \gls{CDS} funding model in
\prettyref{s:cds}.

The \gls{NASA} \gls{PDS} has developed a parameterized model for
helping proposers estimate the costs involved in preparing data for
archiving in the
PDS\footnote{\url{http://pds.nasa.gov/tools/cost-analysis-tool.shtml}};
most relevantly for the above discussion it includes a scaling with
data volume of $1+1.5\log_{10}(\mbox{volume/MB})$ (that is, a
multiplier which increases by 1.5 for each order of magnitude increase
in data volume).


As noted in \prettyref{s:hepdata}, the HEP community is now
constructing more detailed plans for data preservation, and the
associated costs.\label{s:hepdmpplans}  Reference~\cite{south11} estimates that a formal
long-term archive (a level-3 or -4 archive, in the terms of that
paper) would cost 2\range3 FTEs for 2\range3 years after the end of the
experiment, followed by 0.5\range1.0 FTE/year/experiment spent on the
archive's preservation.  They compare this to the 100s of FTEs spent
on for the running of the experiment, and on this basis claim an
archival staff investment of 1\% of the peak staff investment, to
obtain a 5\range10\% increase in output (the latter figure is based on
their estimate that around 5\range10\% of the papers resulting from an
experiment appear in the years immediately after the experiment
finishes; since this latter figure is derived on the current model,
which achieves this without any formal preservation mechanisms, this
estimate of the return on investment in archives may be optimistic).

It is worth noting that in astronomical, HEP and GW contexts, archive
ingest\index{data!ingest} is generally tightly integrated with the
system for day-to-day data management, in the sense that data goes
directly to the archive on acquisition and is retrieved from that
archive by researchers, as part of normal operations.  On the other
side of the archive, projects will generate and disseminate data
products\dash which look very much like OAIS \gls{DIP}s\dash as part of
their interaction with external collaborators, without regarding these
as specifically archival objects.  Thus the submissions into the
archive may consist of both raw data and things which look very much
like DIPs, and the objects disseminated will include either or both
very raw and highly processed data.  The \emph{long-term} planning
represented in the LIGO DMP plan\index{LIGO!DMP}~\cite{anderson11}, for example, is
therefore less concerned with setting up an archive, than with the
adjustments and formalizations required to make an existing
data-management system robust for the archival long term, and more
accessible to a wider constituency.  What this means, in turn, is that
some fraction of the OAIS ingest and dissemination costs (associated
with quality control and metadata, for example) will be covered by
normal operations, with the result that the \emph{marginal} costs of
the additional activity, namely
long-term archival ingest and dissemination,  are probably both rather low
and typically borne by infrastructure budgets rather than requiring
extra effort from researchers.\footnote{This is consistent with the
  ERIM project's conclusions that \qq{ideally information management
    interventions should result in a zero net resource
    increase}~\protect\cite[p.8]{darlington11}.  In this case there is no
  extra resource required from the researchers, though there might be
  a need for extra resource under an infrastructure heading.}
This is corroborated by our
informants above, who generally regard archive costs as coming under a
different heading from \q{data processing costs}.  The point here is
not that the OAIS model does not fit well\dash it fits very well indeed\dash
nor that ingest and dissemination do not have costs, but that if the
associated activities can be contrived to overlap with normal
operations, then the costs directly associated with the archive may be
significantly decreased.  This is the intuition behind the recent
developments in \q{archive-ready} or \q{preservation-aware storage} (cf
\cite{factor07} and \prettyref{s:dcc-lifecycle}), and confirms that it is a viable and effective
approach.

As a final point, we note that big-science projects are inevitably
also large-scale engineering projects, so that the consortia and
their funders are broadly familiar with the procedures, uncertainties
and management of cost estimates, so that the costing and management
of data preservation can be naturally built in to the relationship
between funders and funded, if the funders so require it.

As is shown by the vagueness of some of the remarks above (despite
sometimes very specific numbers), there seems
little in the way of a consensus model for the costing of the
long-term preservation of large-scale data.  There will surely be detailed
costings for the management of PB-scale data for commercial
organisations, but these are not likely to be useful for our purposes,
since they are more concerned with immediate business continuity than
multi-decade archives, are serving different technical communities,
and are likely to be extremely confidential.

We therefore recommend that \recommendation{STFC should develop a
  costings model for the publication and preservation of data, which
  is matched to the data challenges of the big-science community}.
We expect that this can build on the domain-agnostic work
already done in this area by JISC, and on the detailed work done on
closely related problems by \gls{NASA}'s cost-estimation
community~\cite{nasa:ceh08}.
%
%
\index{costs|)}

\subsection{The GW community and the AIDA toolkit}
\label{s:aida}

The AIDA Self-Assessment Toolkit~\cite{aida09}\index{AIDA} is a (\gls{JISC}
funded) set of qualitative benchmarks for discussing at how developed
an institution's archive is.  It leads an archive manager through a
set of a few dozen elements, inviting them to grade their archive from
1 (poor) to 5 (international exemplar).\footnote{The AIDA document
  links these five stages, rather alarmingly, to a five-step programme developed at Cornell,
  which starts with acknowledging that you have a problem, and goes,
  via institutionalisation, to \qq{embracing [\dots] dependencies}, noting that
  \qq{you can't do it alone}.  Clearly, data-management planning is habit-forming.}
The goal is not to produce a
pass/fail score, but instead to help archive managers understand their
current and future requirements, and to \qq{enable an institution to decide
whether specific actions need to be taken in regard to particular
assets, or when and how it is desirable to improve on its current
capabilities}.  The AIDA authors acknowledge that the assessment is
simplistic and subjective, but stress that \qq{AIDA
aims to allow you to evaluate your institution against a recognised
capability scale, and then suggests appropriate actions based on that
evaluation}.  The AIDA goal is to model the progress of an archive
from the acknowledgement that an archive is desirable, through to the
exemplary externalisation of the archive as a resource.

In Appx.~\ref{s:aida-scores}, we list our estimates of the scores for
LSC data management.  We hope these assessments are of specific use to
the GW community, but believe that the discussion in general may be of
use to other, similarly structured, \gls{big science} communities.

The scores for the current LSC cluster in the middle, around three
(which corresponds to \q{consolidate} in the Cornell model).  This is
an impressive score for a project which is, from one point of view,
doing only what is regarded as normal for a well-run large-scale
physics experiment.  The higher scores are generally associated with
the formality and auditability of the long-term plans, rather than any
qualitatively different practice, and we believe that these scores
will naturally drift upwards as a result of the development of an
explicit DMP plan, structured using the OAIS concept set, in collaboration
with a suitably critical funder.


The toolkit is broken into organisational, technology and resources
(generally funding) \q{legs}.

The \q{organisational leg} is concerned with the high-level support
for the archive.  To the extent that it is meaningful, the average for
these scores is above three (which is good).  The lower scores are
generally associated with the informality of the current archive
(compared to a service-oriented commercial organisation) rather than
any more concrete inadequacy: the data is backed up and reasonably
findable, though this reflects cultural norms within the physical
sciences rather than something a particular archiving plan can take
credit for.

The \q{technology leg} is concerned with the hardware and personnel
support for data management.  As with the organisational leg, the GW
community scores highly here without really trying, simply because the
community has long experience of managing \emph{and sharing} large volumes of
data.  The lower scores are again associated with the current informality of
operations (from the point of view of an archive as opposed to a working
data-management infrastructure), and these will naturally rise when
the LSC's \gls{DMP} plan is implemented and reviewed.

The scores in the \q{resources leg} are the least well-justified.  The
LSC generally scores well, in the sense that we can be confident that
there will be resources to support an archive effort\dash it's seen as a
high-importance activity\dash even though there are few
resources currently explicitly earmarked for this.  This section may
therefore be useful for suggesting what budget lines should eventually
exist.

\section{Conclusions and recommendations}

In this report, we have described some of the ways in which \q{big science}
manages its data, as part of a broader data culture which is
characterised by large collaborations, and which has decades of
experience in agreeing how, and when, and when not, to share data.

We can say with some confidence that the big science data culture
manages its data well (and this seems to be corroborated by the AIDA
assessment discussed in \prettyref{s:aida}), but we are not suggesting
that other disciplines could or should simply copy this culture, since
there are various reasons (cf, \prettyref{s:big-science-easy}) why
this culture is particularly natural in some areas.

There are however some practices which we do believe are
straightforwardly portable to other disciplines.  As we discuss in
\prettyref{s:reification}, the notions of \emph{data products} and
\emph{proprietary periods} very naturally concretize otherwise diffuse
arguments about data management and sharing, transforming them from
\q{whether} and \q{why} to \q{which} and \q{how long}.  As well, we
believe that embedding data management in the day-to-day practice of
researchers lowers costs in both the short term (researchers can
easily re-find their own data, and interpret others') and the long
term (since preservation becomes a technical problem of conserving an
in-use repository).  We discuss the costing of data management at
slightly greater length in \prettyref{s:preservation-costs}.

We repeat our explicit recommendations below.
\recommendations

\subsection*{Acknowledgements}

This project was funded by JISC, as part of the \q{Managing Research Data} programme.
We are most grateful to the numerous people who have commented on
various drafts of this report, or provided us with information or resources.
In particular, we thank
Stuart Anderson (LIGO),
Paul Butterworth (NASA),
Harry Collins (Cardiff),
Fernando Comer\oacute n (ESO),
Joy Davidson (DCC),
Fran\ccedilla oise Genova (CDS),
Magdalena Getler (DCC),
Simon Hodson (JISC),
Sarah Jones (DCC),
Dorothea Salo (Wisconsin),
Angus Whyte (DCC),
and
Roy Williams (LIGO).

\appendix

\begingroup
\section{Case study}
\label{s:casestudy}

We have produced a detailed discussion of the structure of the LIGO
working data management system, as a separate
document~\cite{carozzi10}.  This document is currently available only
within LIGO: those observations which have not been incorporated into
this present report are probably too detailed to be of general
interest.  We hope, however, that the case-study will be of some use
internally to to the LSC.

%
%

\endgroup

\section{AIDA assessment}
\label{s:aida-scores}

The AIDA self-assessment toolkit~\cite{aida09}\index{AIDA} is a
\gls{JISC}-funded set of qualitative benchmarks for assessing how developed
an institution's archive is.  See \prettyref{s:aida} for discussion.

The labels in the table below are sometimes a little cryptic; refer to
the full toolkit for useful elaborations.

The answers below generally refer to the \emph{early 2011} state of the
LSC archive arrangements, on the grounds that concrete answers to a
variant question are preferable to speculative answers to a future
one.  These are probably a reasonable indication of the likely status
of a forthcoming formal archive, but in a few case, as noted, we can
give no meaningful answer.

In the scores below, level~1 is \q{poor}, and level~5 is
\q{international examplar}.

\begingroup                     
\makeatletter
\newcommand\aidaelement[5]{%
  \def\@tempa{#4}%
  \ifx\@tempa\@empty
    \item[#1: #3]#5
  \else
    \item[#1: #3 (#4)] #5
  \fi
}
\makeatother

\subsection*{Organisational leg}
\begin{description}
\aidaelement{1}{Mission statement}{institution-wide mission
  statements}{5}{The LIGO project has prepared a formal DMP, at funder request}
\aidaelement{2}{Policies and procedures}{institutional policies for asset management}{3}{LIGO has prepared a formal DMP, and is addressing political and cultural reservations, awaiting funding and implementation}
\aidaelement{3}{Policy review}{review mechanisms at Institutional level}{4}{As well as the DMP, there already exist well-understood collaboration-wide review procedures, and these will be used to review the plan on an annual basis}
\aidaelement{4}{Asset management and sharing}{institutional capability for sharing assets}{3}{Current storage is, of necessity, distributed; the collaboration manages this informally but effectively, however this is generally working storage, and not regarded as archival storage}
\aidaelement{5}{Asset continuity}{institutional level of contingency planning}{3}{There is no formal centralised asset management.  Continuity is regarded as a technical matter which can reasonably be left to the professional good practice of the sites managing the distributed storage.  As before, this is currently regarded as working rather than archival storage.}
\aidaelement{6}{Audit trail of activities and use}{institutional capability for audit}{3}{Extensive logs exist, but are not centralised nor in any standard format; files, once created, are not expected to be modified, though there is no way to verify that this is true in fact}
\aidaelement{7}{Monitoring and feedback from users}{institutional monitoring mechanisms}{4.5}{All data and processes are open to the entire collaboration, and most processes are widely discussed; the collaboration is its own user-base. There are (by design) no external users of the data, nor yet any external review of the mechanisms.}
\aidaelement{8}{Metadata management}{extent of institutional conformance to metadata management}{2 to 4}{Meta\-data is devised in a somewhat \adhoc\ way by individual instruments or software elements (stage 2), but this is also added and managed thoroughly, and in accordance with what is regarded as experimental good practice (stage 4)}
\aidaelement{9}{Contractual agreements}{extent of institutional contracts}{3}{Not applicable to current working storage}
\aidaelement{10}{IPR and rights management}{institutional understanding of IPR}{5}{Formal MoUs between partners regarding access to data, and clear guidance from funders regarding the eventual release of the data}
\aidaelement{11}{Disaster planning and business continuity}{institutional disaster planning}{2}{As with asset continuity, this is currently regarded as a technical matter for storage managers}
\end{description}

\subsection*{Technological leg}
\begin{description}
\aidaelement{1}{Technological infrastructure}{institutional infrastructure}{5}{The collaboration has considerable technical resource, and interoperates well.  Planning is informal but effective.  The sophisticated user-base is comfortable with this informality, but this could in principle become a liability when the resource management moves from a development to a service model.}
\aidaelement{2}{Appropriate technologies}{appropriateness of institutional technologies}{4}{There is plenty of appropriate  technology, though the plan for the archival management of assets is not yet detailed}
\aidaelement{3}{Ensuring availability and integrity}{integrity of institutional backup and storage}{3}{Important data is backed up (possibly by mirroring), as part of normal operations}
\aidaelement{4}{Integrity of information}{institutional processes}{2}{Uncertain: what there is will be done as part of normal operations}
\aidaelement{5}{Obsolescence}{institutional understanding of obsolescence}{3.5}{High general awareness, and occasional discussion, but at present little formal planning}
\aidaelement{6}{Changes to critical processes}{institutional capability}{4}{Changes to processes are widely and frankly discussed, and documented as internal publications; change is managed effectively, but relatively informally}
\aidaelement{7}{Security of environment}{institutional capability for security}{3}{There is a high level of awareness of the need to keep the data proprietary, but given the scientific context, there are no likely attack scenarios as such; the problem will largely evaporate once the data is finally released publicly}
\aidaelement{8}{Security mechanisms}{institutional security mechanisms}{3.5}{No formal threat analyses, but the security is probably appropriate to the level of threat; day-to-day attacks (ie not specifically targeted at this data) are the responsibility of distributed storage and computing managers}
\aidaelement{9}{Implementation of disaster recovery plan}{institutional disaster plan and capacity for business recovery}{3}{Not applicable to the current experimental phase}
\aidaelement{10}{Metadata creation}{institutional capacity to create metadata}{4}{Almost all metadata is added automatically (compare organisational.08)}
\aidaelement{11}{Institutional repository}{effectiveness of an Institution-wide repository}{2}{LIGO has prepared a formal DMP}
\end{description}

\subsection*{Resources leg}
\begin{description}
\aidaelement{1}{Business planning process}{institutional business planning}{2}{LIGO is preparing a formal DMP}
\aidaelement{2}{Review of business plan}{institutional capacity for review}{4}{DMP to be reviewed annually; project as a whole has close relationships with funders and stakeholders}
\aidaelement{3}{Technological resources allocation}{institutional capability for resource allocation}{4}{Resource planning is coordinated at a senior level}
\aidaelement{4}{Risk analysis}{institutional capability for risk management}{2}{General awareness at present, but this should become clearer in future DMP iterations}
\aidaelement{5}{Transparency and auditability}{institutional business transparency}{4.5}{Depending on the precise meaning intended, this could be 4 or 5.  There is substantial auditing from collaboration funders}
\aidaelement{6}{Funding}{institutional capacity for sustainable funding}{3.5}{Good relationships with funders mean that funding is probably predictable on five- to ten-year time\-scales, but unpredictable in the longer term.  However the main funder (NSF) has expressed a strategic commitment to long-term data preservation.}
\aidaelement{7}{Staff skills}{institutional staff management}{3}{Not applicable to the current experimental phase}
\aidaelement{8}{Staff numbers}{institutional management of staff numbers}{3}{Not applicable to the current experimental phase}
\aidaelement{9}{Staff development}{institutional commitment to staff development}{3}{Not applicable to the current experimental phase}
\end{description}

\endgroup

\newpage
\section*{About this document}
\addcontentsline{toc}{section}{About this document}
\label{s:documenthistory}

\begin{description}
\item[LIGO-P1000188-v10, 2011 June 29] First public version, available
  at \url{https://dcc.ligo.org/cgi-bin/DocDB/ShowDocument?docid=p1000188}
\item[v1.1, 2012 July 14] Minor revisions, some added material, and
  typos and minor errors corrected.  There are a couple of additional
  sections, but no changes to section or figure numbers.  The
  pagination will have changed in places.
\end{description}
\ifsnapshot
\begingroup
\parindent0pt
\leftskip=2.5cm
\parskip=0pt
\let\\\medskip
\def\tag#1#2{\noindent\hbox to 0pt{\hss #1\quad}#2\par}
\InputIfFileExists{changelog.tex}{}{}
\endgroup
\fi

\ifendnotes
\clearpage
\begingroup
\parindent 0pt
\parskip 1ex
\def\enotesize{\normalsize}
\def\enoteheading{\section*{Notes}\addcontentsline{toc}{section}{Notes}}
\theendnotes

\endgroup
\fi

\clearpage
\begingroup
\ifxetex
  \def\onedash{-}
  \catcode`\-=\active
  \def-#1{\ifx#1-–\else \onedash #1\fi}
\fi
\bibliography{../mrd-gw}
\bibliographystyle{unsrturl}
\endgroup

\clearpage
\index{LIGO!Advanced|see{glossary: aLIGO}}
\addcontentsline{toc}{section}{Glossary and index}
\def\glossarypreamble{\emph{Terms marked \q{OAIS} are copied from the
    OAIS specification \cite[\S1.7.2]{std:oais}.}}
\def\glsgroupskip{} 
\printglossaries

\def\see#1#2{\emph{see:} #1}
\InputIfFileExists{report.ind}{}{No index found}

\end{document}